\UseRawInputEncoding
\documentclass[a4paper,12pt]{article}
\pdfoutput=1
\usepackage{epsfig}
\usepackage{amssymb}
\usepackage{hyperref}
 \usepackage{color} 
\usepackage{bookmark}

\usepackage{setspace}

\usepackage{amsmath}
\usepackage{amssymb}
\usepackage{graphicx}
\usepackage{geometry}
\usepackage{subfigure}
\usepackage{url}
\usepackage{slashed}
\usepackage{cite}

\newcommand{\ie}{{\it i.e.}}

\newcommand{\be}{\begin{equation}}
\newcommand{\ee}{\end{equation}}
\newcommand{\br}{\begin{eqnarray}}
\newcommand{\bea}{\begin{eqnarray}}
\newcommand{\eea}{\end{eqnarray}}
\newcommand{\er}{\end{eqnarray}}
\newcommand{\ba}{\begin{array}}
\newcommand{\ea}{\end{array}}
\newcommand{\bi}{\begin{itemize}}
\newcommand{\ei}{\end{itemize}}
\newcommand{\bn}{\begin{enumerate}}
\newcommand{\en}{\end{enumerate}}
\newcommand{\bc}{\begin{center}}
\newcommand{\ec}{\end{center}}

\newcommand{\Eq}[1]{Eq.~(\ref{#1})}
\newcommand{\rfn}[1]{(\ref{#1})}

\newcommand{\beq}{\begin{equation}}
\newcommand{\eeq}{\end{equation}}

\newcommand{\U}{\scriptscriptstyle U}
\newcommand{\D}{\scriptscriptstyle D}

\newcommand{\EE}{\scriptscriptstyle E}
\newcommand{\NN}{\scriptscriptstyle N}
\newcommand{\E}{\scriptscriptstyle E}

\newcommand{\DP}{\gamma_{\!\scriptscriptstyle{D}}}
\newcommand{\alphaDP}{\alpha_{\!\scriptscriptstyle{D}}}
\newcommand{\subDP}{\gamma_{\!\raisebox{-2pt}{$\scriptscriptstyle D$}}}

\newcommand{\gsim}{\lower.7ex\hbox{$\;\stackrel{\textstyle>}{\sim}\;$}}
\newcommand{\lsim}{\lower.7ex\hbox{$\;\stackrel{\textstyle<}{\sim}\;$}}

\newcommand{\bs}{\begin{small}}
\newcommand{\es}{\end{small}}

\newcommand{\qui}{q_{{\scriptscriptstyle U}_{\!i}}}
\newcommand{\qdi}{q_{{\scriptscriptstyle D}_{\!i}}}
\newcommand{\qei}{q_{{\scriptscriptstyle E}_{\!i}}}
\newcommand{\qni}{q_{{\scriptscriptstyle N}_{\!i}}}
\newcommand{\eu}{e_{{\scriptscriptstyle U}}}
\newcommand{\ed}{e_{{\scriptscriptstyle D}}}

\newcommand{\BR}{{\rm BR}}
\def\G{\scriptscriptstyle{G}}

\begin{document}
\pagestyle{empty}
\begin{center}
{\Large {\bf 
    Dark photon searches via Higgs boson production  
\vspace{0.5cm}
  \\ at the  LHC and beyond
}} \\
\vspace*{1.5cm}
{
  {\bf Sanjoy Biswas$^{a}$}, 
 {\bf Emidio Gabrielli$^{{b,c}}$},   
 {\bf Barbara Mele$^{d}$}
}\\

\vspace{0.5cm}
       {\it
    (a)   School of Mathematical Sciences,  
Ramkrishna Mission Vivekananda Educational and Research Institute,  
P.O. - Belur Math, Howrah - 711202, India}
\\[1mm]
{\it  (b) Dipart. di Fisica Teorica, Universit\`a di 
Trieste, Strada Costiera 11, I-34151 Trieste, Italy and 
INFN, Sezione di Trieste, Via Valerio 2, I-34127 Trieste, Italy}  
\\[1mm]
{\it
  (c) NICPB, R\"avala 10, 10143 Tallinn, Estonia}  \\[1mm]
 {\it
   (d) INFN, Sezione di Roma, c/o Dipartimento di Fisica, Sapienza Universit\`a di Roma, \\ Piazzale Aldo Moro 2, I-00185 Rome, Italy}  \\[1mm]

\vspace*{2cm}{\bf ABSTRACT}
\end{center}

\vspace{0.3cm}

\vspace*{5mm}
Many scenarios beyond the standard model, aiming to solve long-standing cosmological and particle physics problems, suggest that dark matter might experience long-distance interactions mediated by an {\it unbroken} dark $U(1)$ gauge symmetry, hence foreseeing the existence of a {\it massless} dark photon. Contrary to the massive dark photon, a {\it massless} dark photon can only couple to the standard model sector by means of effective higher dimensional operators.  Massless dark-photon production at colliders will then in general be suppressed at low energy by a UV energy scale, which is of the order of the masses of portal (messenger) fields connecting the dark and the observable sectors. A violation of this expectation is provided by  dark-photon production mediated by the Higgs boson,  thanks to the non-decoupling Higgs properties. Higgs-boson production at colliders, followed by the Higgs decay into a photon and a dark photon, provides then a very promising production mechanism for the dark photon discovery, being insensitive in particular regimes to the UV scale of the new physics. This decay channel gives rise to a peculiar  signature characterized by a monochromatic photon with energy half the Higgs mass (in the Higgs rest frame) plus missing energy. We show how such resonant photon-plus-missing-energy signature can {\it uniquely} be connected to  a dark photon production.  Higgs boson production and decay into a photon and a dark photon  as a source of dark photons is reviewed at the Large Hadron Collider, in the light of the present bounds on the corresponding signature by the CMS and ATLAS collaborations. Perspectives for the dark-photon production  in Higgs-mediated processes at  future $e^+e^-$ colliders are also discussed.
\noindent

\vfill\eject

\pagestyle{plain}

\section{Introduction}

The discovery of the Higgs boson in 2012 at the Large Hadron Collider (LHC) by the  {\mbox{ATLAS}}~\cite{ATLAS:2012yve} and CMS~\cite{CMS:2012qbp} collaborations has been a milestone for particle physics and the triumph of the Standard Model (SM) of electroweak (EW) and strong interactions \cite{CMS:2022dwd,ATLAS:2022vkf}. The good experimental agreement with the SM Higgs expectations has strengthened our confidence in the Higgs mechanism and in the existence of the Yukawa couplings to fermions, needed for the fermion mass generation in the SM framework \cite{ATLAS:2016neq,ATLAS:2019slw}. Recent results
from the  LHC data analysis have further consolidated these expectations 
\cite{CMS:2018uag,Sirunyan:2018kst,Aad:2020vbr,Aaboud:2018zhk,Sirunyan:2017khh,ATLAS:2018lur,Sirunyan:2018hoz,Aaboud:2018urx}.
In particular, the observations of the Higgs boson decay modes into bottom-quark~\cite{Sirunyan:2018kst,Aad:2020vbr,Aaboud:2018zhk} and tau-lepton pairs~\cite{Sirunyan:2017khh, ATLAS:2018lur}, as well as the detection of the Higgs boson production in association with top-quark pairs  \cite{Sirunyan:2018hoz, Aaboud:2018urx}, are all consistent with the hypothesis of a SM Yukawa coupling strength, thus supporting the existence of the corresponding interactions in Nature.

Despite the good SM agreement with data, we are still far from a complete understanding of the Higgs boson physics and of its properties. Regardless of the Higgs boson discovery, a few major intriguing puzzles, highlighted below, are still to be clarified.
\begin{itemize}
  \item
    According to the common wisdom,  New Physics (NP) at the TeV scale, charged under the SM interactions, is needed to stabilize the electroweak vacuum against potentially large quantum corrections to the Higgs potential, often referred to as the {\it fine-tuning} problem (see for instance \cite{Barbieri:1987fn,Barbieri:2006dq,Giudice:2008bi}). So far, no such NP has been discovered at the LHC, raising doubts about our understanding of {\it naturalness} \cite{tHooft:1979rat} in quantum field theory.
\item
  The SM does not contain  suitable dark matter candidates. Although dark matter is five times more  abundant than ordinary baryonic matter \cite{Planck:2015fie,Bertone:2016nfn,Simon:2019nxf,Salucci:2018hqu,Allen:2011zs,Bahcall:2000zu}, its properties are yet unknown.  
\item
An underlying mechanism   explaining the origin of the large hierarchy among fundamental fermion masses, or analogously their Yukawa couplings, including the origin of the small neutrino masses and flavor mixing, is missing in the SM.
\end{itemize}

While the fine-tuning issue might be an ill-defined problem, as recently emphasized in \cite{Mooij:2021ojy}, the presence of dark matter in the universe is a real experimental evidence for NP \cite{Planck:2015fie,Bertone:2016nfn,Simon:2019nxf,Salucci:2018hqu,Allen:2011zs,Bahcall:2000zu}, in case its nature is explained by the presence of new weakly interacting constituents of non-baryonic origin beyond neutrinos, which are missing in the SM spectrum.

Despite its growing evidences, the non-gravitational nature of  dark matter remains a mistery, so far eluding all {\it direct}~\cite{Bernal:2017kxu,Bergstrom:2000pn,Bertone:2016nfn,deSwart:2017heh,Arcadi:2017kky} and {\it indirect} (\ie\ based on the search for  annihilation or decay debris of hypothetical dark matter particles~\cite{Khlopov:2014nva,Gaskins:2016cha}) detection.
This factual observation has recently opened the way to more speculative approaches about its origin. One intriguing possibility is that dark matter could be linked to the presence of a \textit{dark sector} beyond the SM\cite{DiValentino:2019jae,Buen-Abad:2017gxg,Kumar:2017dnp,Cirelli:2016rnw,Berlin:2016vnh,Elor:2015bho,Foot:2014uba,Berlin:2014pya,Boddy:2014yra,Khlopov:2013ava,Cohen:2010kn,Bean:2008ac}. A dark sector is made of states that are singlets under the SM gauge groups, and  can also have its own structure and interactions. Indeed, dark matter might be even charged under its own long-range force (that is not experienced by SM particles), mediated for instance by a hidden $U(1)$ gauge symmetry in the dark sector \cite{Ackerman:2008kmp}. Speculative approaches in this direction have been motivated in parts by potential discrepancies in conventional dark matter scenarios (especially on small scales), but also by the fact that {\it charged} dark matter could help  explaining galaxy formation and dynamics \cite{Arkani-Hamed:2008hhe,Fan:2013tia,Heikinheimo:2015kra,Agrawal:2016quu}.

The interest for dark sector searches goes anyhow beyond the purposes of astrophysics and cosmology, as shown by a number of recent reports on the subject~\cite{Proceedings:2012ulb,Essig:2013lka,Raggi:2015yfk,Deliyergiyev:2015oxa,Alekhin:2015byh,Alexander:2016aln,Beacham:2019nyx,Albert:2022xla,Boveia:2022jox,Rizzo:2022qan}, covering also searches for long-lived  particles~\cite{Alimena:2019zri} and millicharged particles~\cite{Badertscher:2006fm,Hagley:1994zz,Prinz:1998ua,Magill:2018tbb}.
New phenomenological evidences supporting the possible existence of a dark sector are growing. Recently, a new excess in the electronic recoil data was observed in the XENON1T detector, that could be explained by the presence of a dark photon\cite{Holdom,Fayet:1990wx,Fayet:1980ad,Fayet:1980rr,Okun:1982xi,Georgi:1983sy}, associated to the quanta of a $U(1)$ long-range force in the dark sector, with mass of the order of 2-3 keV \cite{Aprile:2020tmw,Proceedings:2022hmu}.

From the side of flavor physics, all quark-flavor and CP-violating experiments over the last 40 years have confirmed the correctness of the SM description via the Higgs Yukawa interactions for fundamental fermions \cite{Buchalla:2008jp,Antonelli:2009ws}. On the other hand, the large fermion-masses  (or, analogously, Yukawa-couplings) spectrum, which spans over 6 orders of magnitude for charged fermions (and even more if neutrinos have only Dirac masses just as quarks and charged leptons) still remains a mystery.
The lack of any mechanism in the SM  to naturally explain this hierarchy might well suggest the presence of NP behind it. Although the discovery of non-vanishing neutrino masses can hint at a new intermediate scale between the weak and the Planck scale via the sea-saw mechanism, the latter cannot also explain the large gap in the charged-fermion sector.

A new paradigm for the flavor genesis \cite{Gabrielli:2013jka,Gabrielli:2016vbb} has been recently proposed suggesting that SM fermion masses, flavor mixing \cite{Gabrielli:2019sjg}, and dark matter constituents might have a common origin in a dark sector. If correct, this paradigm can provide the missing link between some of the long standing puzzles in particle physics and the existence of dark sectors. The key idea is based on the assumption that the Yukawa couplings are not fundamental couplings, but rather effective ones, radiatively induced by a dark sector. The absence of any tree-level Yukawa operator is guaranteed by some local or global symmetry which eventually is spontaneously broken by the vacuum expectation value (vev) $\mu$ of some scalar field. Contrary to the SM case, for energies above the $\mu$ scale, the Yukawa operators cease to exist as local operators.\footnote{The mechanism to generate effective Yukawa couplings aiming to explain the hierarchy problem has been previously considered. Known attempts in the literature are mainly based on the Nielsen-Froggatt mechanism \cite{Froggatt:1978nt} or confining different fermions in different branes located in different places in extra dimensions \cite{Arkani-Hamed:1999ylh}. However, these mechanisms are both affected by the presence of non-renormalizable interactions (in 4-dimensions), while the approach in \cite{Gabrielli:2013jka} is based on a fully renormalizable theory in 4-dimensions.}

In order to radiatively generate Yukawa couplings at one loop from a dark sector, a set of messengers fields -- charged under SM gauge interactions and with same quantum numbers as squarks and sleptons in supersymmetric (SUSY) models -- 
 is required in order to communicate interactions between the dark and SM observable sectors.  Massive vector-like dark fermion fields, heavier SM gauge-singlet replicas of the SM fermions, are also needed \cite{Gabrielli:2013jka}, with  their masses playing the role of the chiral symmetry breaking parameters. Thus, dark fermions turn out to be almost a heavy replica of the SM fermions, provided the messenger sector is flavor blind. All dark fermions, as well as messenger fields, are also charged under an unbroken $U(1)_{\!\scriptscriptstyle{D}}$ dark-gauge interaction~\footnote{The $U(1)_{\!\scriptscriptstyle{D}}$ gauge symmetry was introduced in \cite{Gabrielli:2013jka} for dynamical reasons. Indeed, it was shown that a particular non-perturbative dynamics in the  $U(1)_{\!\scriptscriptstyle{D}}$ sector can generate an exponentially suppressed spectra of dark-fermions, as a function of the $U(1)_{\!\scriptscriptstyle{D}}$ quantum charges (that are parameters of order ${\cal O}(1)$), thus naturally solving the flavor hierarchy problem according to the naturalness criteria of \cite{tHooft:1979rat}.  See \cite{Gabrielli:2013jka,Gabrielli:2007cp} for more details.}, which automatically ensures dark-fermions stability. The presence of long-range interacting multi-dark matter constituents comes out as one of the main features of this scenario.

It was also recently shown \cite{Gabrielli:2021lkw} that the presence of a radiative Yukawa coupling for the top quark in this framework can fully stabilize the Higgs scalar potential, naturally solving the problem of the vacuum instability of the SM Higgs sector~\cite{Bezrukov:2009db,Ellis:2009tp,EliasMiro:2011aa,Isidori:2007vm,Mihaila:2012fm,Degrassi:2012ry,Chetyrkin:2012rz} without a particular tuning of the parameters.

From considerations based on general grounds, the SM  fields can couple to the dark-sector fields by means of higher dimensional operators, whose associated effective scale $\Lambda$ is expected to be proportional to the average mass of the corresponding messenger sector. Therefore, the sensitivity to the dark-sector searches at  low energies $E$ is expected to be suppressed by some powers of $E/\Lambda$, that depend on the dimension of the operator involved.

A deviation from this rule is provided by the coupling of the dark photon, the quanta associated to the corresponding field of the $U(1)_{\!\scriptscriptstyle{D}}$ gauge interaction in the dark sector. 
The dark photon scenario \cite{Holdom,Fayet:1990wx,Fayet:1980ad,Fayet:1980rr,Okun:1982xi,Georgi:1983sy} has been extensively analyzed in the literature, and  has also been the subject of many current experimental searches \cite{Alexander:2016aln} (see \cite{Fabbrichesi:2020wbt} for a recent review on the dark photon physics). 

Indeed, a dark photon can have  renormalizable couplings to SM fields\cite{Holdom}, induced by a tree-level kinetic mixing with an ordinary photon, namely $\epsilon \, F_{\mu\nu}{F}_{\mu\nu}^{\scriptscriptstyle{D}}$, where $F_{\mu\nu}$ and ${F}_{\mu\nu}^{\scriptscriptstyle{D}}$ are the field strengths of the photon and dark photon, respectively. The dimensionless mixing  parameter $\epsilon$ is expected to depend only logarithmically on the UV scale. After diagonalizing the kinetic term, a massive dark photon can acquire a millicharged tree-level coupling to an ordinary SM charged particle, which is  proportional to the $\epsilon$ parameter~\cite{Holdom}.  For dark-photon  masses above 1 MeV, the dark photon can then decay into SM charged leptons.
Consequently, strong limits on the $\epsilon$ parameter for dark-photon  masses above 1 MeV have been set from direct di-lepton searches at colliders and fixed target experiments, as well as indirectly from supernovae~\cite{Fabbrichesi:2020wbt}.

On the other hand, for a massless dark photon things go in a different way. The kinetic mixing can be fully rotated away, and  no tree-level coupling with charged SM fields is left. At the same time, a massless dark photon can potentially acquire couplings to SM fields via higher-dimensional operators~\cite{Holdom}. This would make in general a massless dark-photon search different from the massive case, and strongly dependent on the size of the effective scale of the associated operator.
On the other hand, the fine structure constant 
($\alphaDP$) characterizing the dark $U(1)_{\!\scriptscriptstyle{D}}$
interaction can be relatively large, presently missing relevant experimental 
constraints~\cite{Fabbrichesi:2020wbt}.

In this paper,  we restrict our discussion of the dark-photon production to  Higgs boson mediated processes. The latter manifests crucial {\it non-decoupling} properties, that makes the corresponding rates insensitive to the UV effective scale, and controlled by the electroweak scale, just as in the one-loop 
photon-photon ($H\gamma\gamma$) or gluon-gluon ($Hgg$) Higgs SM amplitudes (for a review on more general Higgs couplings to the dark sector see~\cite{Lagouri:2022oxv}).

Indeed, both massless and massive dark photons can have effective couplings with the Higgs boson induced by the exchange of  messenger fields~\cite{Gabrielli:2013jka,Dobrescu:2004wz}. For the Higgs coupling  to a photon and a dark photon ($H\gamma \DP$), the effective coupling is dominated by the  
dimension-5 operator  $H F_{\mu\nu}{F}_{\mu\nu}^{\scriptscriptstyle{D}}$.  Remarkably, as observed in \cite{Gabrielli:2013jka}, this interaction manifests non-decoupling effects with respect to the UV theory, in both massless and massive dark-photon scenarios. In the massive  case   the dark photon can also couple to the Higgs via top-quark and $W^{\pm}$ boson loops through millicharge effects (proportional to the $\epsilon$ parameter). On the other hand, 
the latter  tends to be very much suppressed with respect  to heavy-messenger loop-induced contributions due to the present strong $\epsilon$ bounds for a light dark photon~\cite{Fabbrichesi:2020wbt}.
Hence, dark-photon production mediated by the Higgs boson via messenger 
loops is expected to be the leading dark-photon production channel in both the massless and massive cases.
 
Using the minimal model of \cite{Gabrielli:2013jka} as a (renormalizable) benchmark scenario, one can indeed show that the associated scale depends only on the Higgs vev, and it is independent on the UV scale set by the mass of the (heavy) messenger fields running in the loop.  In particular, for a minimal scenario, this scale will depend only on two free dimensionless parameters, $\alphaDP$, and $\xi$ -- the mixing parameter in the scalar messenger sector  defined in the following-- which satisfies the condition $|\xi|<1$.

The  effective Higgs-dark photon couplings  can induce the visible Higgs decay into a photon plus a dark photon $H\to \gamma \DP$. Since a massless $\DP$ is not detected, the corresponding signature at the LHC would be characterized by an almost monochromatic photon, with energy half of the Higgs boson mass in its rest frame, plus missing transverse energy \cite{Gabrielli:2014oya,Biswas:2016jsh}. Due to the expected non-decoupling properties, this decay might have measurable rates and could be a golden channel for the massless dark-photon discovery. Indeed, other dark-photon production mechanisms at colliders are sensitive to the UV effective scale, being  massless dark photons
coupled  in general via higher-dimensional operators which do experience decoupling properties.
Similar conclusions hold also for the Higgs coupling to two dark photons ($H\DP \DP$), which gives contribution to the invisible width of the Higgs boson.

The above features characterize also
a massive but light dark photon, which is not decaying into visible particles inside the detector.
  
Experimental searches for the mono-photon Higgs signature have been recently carried out at the LHC by the ATLAS \cite{ATLAS:2021pdg} and CMS \cite{CMS:2020krr,CMS:2019ajt} collaborations, for the most promising Higgs production channels of vector boson fusion (VBF), and associated vector boson production (VH). The negative result of these searches have been turned into a few percent upper bounds on the corresponding Higgs $H\to \gamma \DP$ branching ratio (BR).

In this review, we will focus on the physics of the massless (or lightly massive) {\it invisible} dark photon, and its implications for the  $H\to \gamma \DP$ decay rates. We will analyze  the main discovery signatures for this decay channel at the  LHC \cite{Gabrielli:2014oya,Biswas:2016jsh} and future $e^+e^-$ colliders \cite{Biswas:2015sha,Biswas:2017lyg}, including a discussion on  recent relevant ATLAS and CMS analyses~\cite{ATLAS:2021pdg,CMS:2020krr,CMS:2019ajt}.

With respect to previous reviews on  dark photon phenomenology~\cite{Fabbrichesi:2020wbt,Lagouri:2022oxv}, a few relevant theoretical aspects underlying  the dark photon connection to the Higgs boson sector have been scrutinized  and discussed in details. In particular, the relevant effective couplings of the Higgs boson to dark photon and photon (or Z boson) are explicitly shown at one loop by means of a simplified model for the dark sector. We also  provide predictions for the corresponding new physics contributions to the decay channels $H\to \gamma \gamma$, and $H\to 2$\,{\it gluons}. This will allow a model independent analysis  of the $H\to \gamma \DP$ rate, satisfying all constraints arising from  the LHC measurements of Higgs properties. Finally, based on general theoretical arguments, we will show how the detection of a hypothetical $H\to \gamma \DP$ signature could set strong evidence  for the spin-1 nature of the dark photon, definitely disfavouring different-spin candidates  giving rise to missing energy. We will also include a detailed analysis of the dark photon production mediated by the Higgs boson at  future $e^+e^-$ colliders, which has not been covered in~\cite{Lagouri:2022oxv}.

  The paper is organized as follows. In section 2, we  provide the theoretical description of leading Higgs couplings to a massless dark photon, and the results for the corresponding Higgs decay BR's. We also stress the unique connection  between  a two-body $H\to \gamma \, X_{inv}$ signature and the production of a spin-1 (dark photon) particle. In section 3, we  analyze the signal arising from the Higgs decay  $H\to \gamma \DP$ at the LHC in the  main Higgs production channels of gluon-gluon fusion and VBF, discussing also  the dominant  backgrounds. In section 4, most recent LHC experimental results on the $H\to \gamma \DP$ searches by ATLAS and CMS collaborations are presented. Future perspectives for dark photon searches  at future/possible LHC upgrades are also presented. In section 5,   Higgs-mediated dark-photon
production channels at future $e^+e^-$ colliders are discussed, while our conclusions are given in section~6.

\section{Theoretical Framework}
The dark photon is the quanta associated to an abelian $U(1)_{\!\scriptscriptstyle{D}}$ gauge symmetry of a hypothetical dark sector made up of  particles completely neutral under the SM interactions. We will see there are actually two kinds of dark photons -- massless or massive -- whose  theoretical characteristics as well as  experimental signatures can be quite distinct. 

Let us start by reviewing how the dark photon can couple with ordinary matter and gauge fields.
There are different ways a dark photon can communicate with the ordinary world. The most known portal is provided assuming the existence of a tree-level {\it kinetic-mixing} term with ordinary photons in the Lagrangian, namely a term proportional to $\epsilon \, F_{\mu\nu}{F}^{\scriptscriptstyle{D}}_{\mu\nu}$, where $F_{\mu\nu}$ and ${F}_{\mu\nu}^{\scriptscriptstyle{D}}$ are the  field strength of photon and $U(1)_{\!\scriptscriptstyle{D}}$ gauge field, respectively, being $\epsilon$ a small dimensionless parameter.

The physics arising from the kinetic mixing differs for  a truly massless dark photon with respect to a  massive one, regardless of the latter mass size. For   massless dark photons, one can rotate the fields in such a way that the dark photon gets  coupled at tree level only to the dark charged sector, while  
the dark charged matter fields
 acquire also a ``milli" charge proportional to $\epsilon \,{e}_{\!\scriptscriptstyle{D}}$, where ${e}_{\!\scriptscriptstyle{D}}$ is the $U(1)_{\!\scriptscriptstyle{D}}$ charge unit, and correspondingly mildly couple to  ordinary photon as well.  On the other hand, in the massive case, the freedom of field rotation is prevented by the presence of the dark photon mass term in the Lagrangian~\cite{Holdom,Fabbrichesi:2020wbt}. Then, the massive dark photon field in general couples to both the $U(1)_{\!\scriptscriptstyle{D}}$ and SM electromagnetic currents (in the latter case with a ``milli" charge coupling proportional to  $\epsilon \,e$).

A different type of portal  assumes the existence of (typically scalar or fermion) heavy  messenger fields,  that are charged under both the SM and the $U(1)_{\!\scriptscriptstyle{D}}$ gauge sectors. The presence of a tree-level kinetic mixing at any scale is unavoidable in the presence of messenger fields. Indeed, even if a {\it tree-level}  mixing term is assumed to vanish at some high energy scale, the radiative  corrections could regenerate it at low energy scales.
However, in the presence of messenger fields, the massless dark photon can  acquire couplings to ordinary SM particles as well, via higher dimensional operators that can be induced via loop effects.

In conclusion, following the above considerations, the physics of massless and massive dark photons can be summarized as follows :
\begin{itemize}
\item[-] a \underline{massless dark photon} does not couple at tree level to any of the SM  currents and interacts  instead  with ordinary matter only through operators of dimension higher than four;
\item[-] a  \underline{massive dark photon},
 in addition to higher dimensional operators as in the massless case, can
 couple to ordinary matter through a current (with arbitrarily small charge)  via a renormalizable operator of dimension four. The massless limit of this case does not correspond to the massless case above.
\end{itemize}
Because of their different coupling to SM particles, they are characterized by a different phenomenology.

Now we will focus on the phenomenology of a massless dark photon coupled to the Higgs field. The corresponding results can be easily generalized to a  massive dark photon.
\subsection{The model}
As benchmark model for the portal sector, we consider  the scenario 
discussed in~\cite{Gabrielli:2014oya} for the radiative generation of the SM Yukawa couplings. 
The model assumes a generic messenger sector, consisting of {\it left}-doublets (indicated with a ``hat'') and {\it right}-singlets of the $SU(2)_L$ gauge group, namely  $\hat{S}_L^{U_i}$, $\hat{S}_L^{D_i}$ and $S_R^{U_i}$, $S_R^{D_i}$ scalars, respectively, for the colored messengers, and analogous ones for the electroweak messengers $\hat{S}_L^{E_i}$, $\hat{S}_L^{N_i}$ and $S_R^{E_i}$, $S_R^{N_i}$, with a flavour universal mass term for each $i$, with generation index $i=1,2,3$. Due to the fact that all messenger fields in \cite{Gabrielli:2014oya} have universal Yukawa couplings to dark fermions and quarks/leptons (in order to radiatively generate the SM Yukawa couplings), they incidentally have same SM quantum numbers (QN) as squarks and sleptons of  (SUSY)  models. Moreover, due to the fact that dark fermions are charged under $U(1)_{\!\scriptscriptstyle{D}}$, the messengers must carry also additional $U(1)_{\!\scriptscriptstyle{D}}$ charges. 
In Table~\ref{tab1} we report the corresponding QN for colored and EW messenger fields as given in \cite{Gabrielli:2014oya}.
\begin{table} \begin{center}    
\begin{tabular}{|c||c|c|c|c|}
\hline 
{\bf Fields} 
& ${\bf SU(2)_L}$ repr. 
& ${\bf U(1)_Y}$ charge
& ${\bf SU(3)_c}$ repr.
& ${\bf U(1)_{\!\scriptscriptstyle{D}}}$ charge
\\ \hline 
$\hat{S}_L^{\D_i}$
& 2
& 1/3
& 3
& $\qdi$
\\ \hline
$\hat{S}_L^{\U_i}$
& 2
& 1/3
& 3
& $\qui$
\\ \hline
$S_R^{\D_i}$
& 1
& -2/3
& 3
& $\qdi$
\\ \hline
$S_R^{\U_i}$
& 1
& 4/3
& 3
& $\qui$
\\ \hline 
$\hat{S}_L^{\EE_i}$
& 2
& -1
& 1
& $\qei$
\\ \hline
$\hat{S}_L^{\NN_i}$
& 2
& -1
& 1
& $\qni$
\\ \hline
$S_R^{\EE_i}$
& 1
& -1
& 1
& $\qei$
\\ \hline
$S_R^{\NN_i}$
& 1
& 0
& 1
& $\qni$
\\ \hline
$S_0$
& 1
& 0
& 1
& 0
\\ \hline \end{tabular} 
\caption[]{\small 
  Gauge quantum numbers for the strongly-interacting ($S^{\D_i,\U_i}$) and EW color singlet ($S^{\EE_i,\NN_i}$)  messenger fields, and scalar singlet $S_0$, with the $i=1,2,3$ index associated to the SM fermion generations. $U(1)_{\!\scriptscriptstyle{D}}$ is the gauge symmetry  in the dark sector. The electric charge (in  $e$ units) of each field is given by $Q=I_3+\frac{Y}{2}$, where $Y$ is the hypercharge and $I_3$ is the eigenvalue of the third weak isospin generator, while the dark $U(1)_{\!\scriptscriptstyle{D}}$ charges  are in units of the fundamental dark charge ${e}_{\!\scriptscriptstyle{D}}$.
}
\label{tab1}
\end{center} \end{table}

Since we are interested in providing a minimal UV completion for the radiative generation of the effective Higgs boson couplings involving both dark photons and SM gauge bosons, we restrict here only to the interaction of messenger fields with couplings to the Higgs boson~\cite{Gabrielli:2014oya}.
In particular, for the colored messengers sector (omitting the flavor and color indices) the interaction Lagrangian is
\bea
{\cal L}^I_{MS} &=&
\lambda_S S_0 \left(\tilde{H}^{\dag} S^{\U}_L S^{\U}_R+ H^{\dag} S^{\D}_L S^{\D}_R\right)
+ h.c. ,
\label{LagMS}
\eea
where $\lambda$ is a universal coupling, {the doublet messenger fields  components are $S^{\U}_L=(S^{\U_1}_L,S^{\U_2}_L),~ S^{\D}_L=(S^{\D_1}_L,S^{\D_2}_L)$}, and $S_0$ is a singlet scalar field that has a vev.

After the singlet $S_0$ scalar gets a vev $\langle S_0 \rangle$, trilinear Higgs couplings to messenger fields are generated, and effective couplings of the Higgs to dark photons, $H\gamma\, \DP$ and $H \DP \DP$, are induced at 1-loop, and are proportional to the parameter $\mu_S\equiv \lambda_S\langle S_0 \rangle $.
However, after the electroweak symmetry breaking (EWSB), a mixing mass term in the left and right messenger sectors  arises, which is proportional to $\mu_S v$, being $v$ the Higgs vev.

Then, focussing on the left and right messenger fields components,  the free kinetic Lagrangian for a generic $S_{L,R}$ (for each $U$ and $D$ messenger sectors, and  for the EW  sector as well) is
\bea
    {\cal L}^0_{S}&=& \partial_{\mu} \hat{S}^{\dag} \partial^{\mu}\hat{S} - \hat{S}^{\dag} M^2_S \hat{S},
\label{LMS0}
\eea
where  $\hat{S}=(S_L,S_R)$ ({omitting both $U$, $D$ and flavor indices, and
also $SU(2)_L$ indices}), and the  mass term is given by
\begin{equation}
M^2_S = \left (
\begin{array}{cc}
m^2_L & \Delta \\
\Delta & m^2_R 
\end{array}
\right),
\label{M2}
\end{equation}
with $\Delta=\mu_S v$ parametrising the scalar left-right mixing. It is understood that each term inside Eq.~(\ref{M2}) is proportional to the $3\times 3$ unity matrix in the flavor space. According to  the minimal flavor violation hypothesis~\cite{DAmbrosio:2002vsn}, flavour universality for the 
$m_L^2$ and $m_R^2$ mass terms  is assumed. Then, 
for each flavor sector, the $2\times 2$ matrix of Eq.~(\ref{M2}) can be diagonalized by the matrix
\begin{equation}
U = \left (
\begin{array}{cc}
\cos{\theta} & \sin{\theta} \\
-\sin{\theta} & \cos{\theta} 
\end{array}
\right),
\label{UM}
\end{equation}
where $\tan{\theta}=2\Delta/(m_L^2-m_R^2)$, with mass eingevalues
$m_{\pm}^2=\bar{m}^2\pm \Delta/\sin{2\theta}$, and $\bar{m}^2=(m_L^2+m^2_R)/2$ being the average mass squared.

Concerning the dark-photon interaction with the messenger fields, it can be simply obtained  by substituting the partial derivative $\partial_{\mu}$ with the covariant
  derivative $D^{\mu}=\partial^{\mu}+ie_D q A^{\mu}_{\D}$ in the kinetic term of messenger fields in Eq.~(\ref{LMS0}), where $A^{\mu}_{\D}$ is the dark photon field, $e_D$ stands for the unit of $U(1)_D$ charge, and $q$ is the corresponding charge eigenvalue of the field to which the covariant derivative applies. Notice that, after rotating the messenger fields to the corresponding mass eigenstates basis, the interaction   Lagrangian  (${\cal L}^{\rm DP}_{S}$) involving messenger fields and  dark photon remains diagonal in the mass eigenstate basis. Indeed, the messenger fields  $\hat{S}=(S_L,S_R)$ subject to the rotation have the same $U(1)_D$ charge, namely
  \bea
{\cal L}^{\rm DP}_{S}&=& (D_{\mu} \hat{S}_{\rm M})^{\dag} (D^{\mu}\hat{S}_{\rm M})\, ,
  \eea
where $\hat{S}_{\rm M} =(S_1,S_2)$ symbolically indicates the messenger  mass eigenstates (with the $U$, $D$ and flavor indices omitted).

Finally, notice that since all messenger fields are charged under $U(1)_D$ interactions, no mixing among the Higgs field and electroweak messenger fields can arise [assuming  $U(1)_D$ is unbroken] due to $U(1)_D$ gauge invariance or charge conservation.

\subsection{The Higgs decay $H\to \gamma \DP$}
\begin{figure}[t!]
\begin{center}
\includegraphics[width=6in]{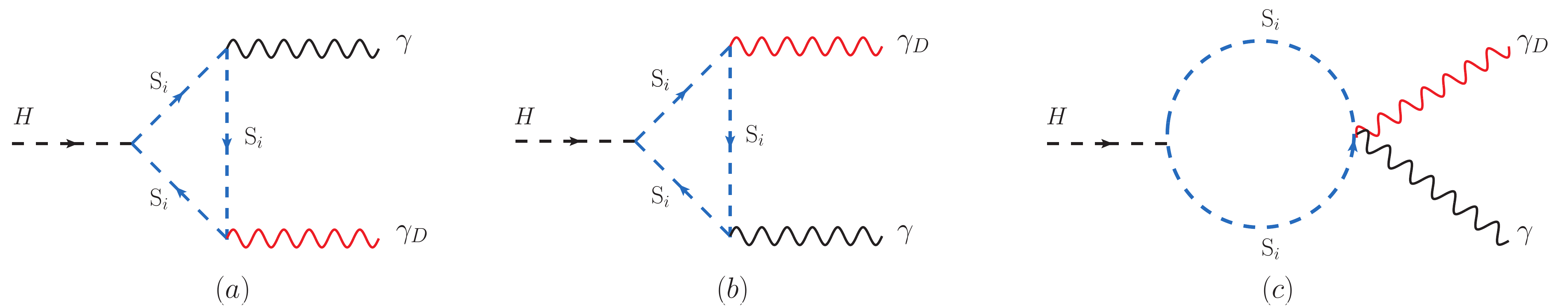} 
\caption{{\small Feynman diagrams contributing to the $H\to \gamma \DP$ decay amplitude, where $\gamma$ and $\DP$ are the photon and dark photon, respectively. The dashed-blue lines  stand for the mass eigenstates  ($S_i$) of messenger scalar fields running in the loop.}}
\label{diagrams}
\end{center}
\end{figure}

After the EWSB, the interaction in Eq.~(\ref{LagMS})  can generate the Higgs boson  decay into a photon  plus a dark photon 
\bea
H\to \gamma \DP\, ,
\eea
whose Feynman diagrams are reported in Fig.~\ref{diagrams}.

If we define  $\varepsilon_1^{\mu}(k_1)$ and $\varepsilon_2^{\mu}(k_2)$ 
the photon and dark photon polarization vectors, respectively,
we can express the $H\to \gamma\, \DP$ amplitude   as 
\bea
M_{\gamma\subDP}&=& \frac{1}{\Lambda_{\gamma\subDP} }\, T_{\mu\nu}(k_1,k_2) 
\varepsilon_1^{\mu}(k_1) \varepsilon_2^{\nu}(k_2),
\label{Mgg}
\eea
where $\Lambda_{\gamma\subDP}$ parametrizes the effective scale associated
to the NP, and the tensor $T^{\mu\nu}$ is given by
\bea
T^{\mu\nu}(k_1,k_2)\equiv g^{\mu\nu} k_1\cdot k_2 -k_2^{\mu}k_1^{\nu}\, ,
\eea
where $k_1$ and $k_2$ are the photon and dark photon  4-momenta, respectively, satisfying the on-shell conditions $k_1^2=k_2^2=0$.
It is easy to verify that the $M_{\gamma\subDP}$ amplitude is gauge invariant, due to the Ward identities $k_1^{\mu}T_{\mu\nu}(k_1,k_2)=k_2^{\nu}T_{\mu\nu}(k_1,k_2)=0$. The total width is then
\bea
\Gamma(H\to \gamma \subDP)=\frac{m_H^3}{32\, \pi\, 
\Lambda_{\gamma \subDP}^2}\, ,
\eea
with $m_H$ the Higgs boson mass.
In order to compute the  $\Lambda_{\gamma \subDP}$ scale, we compute the Feynman diagrams in Fig.~\ref{diagrams}, and match the resulting amplitude  with the expression in Eq.~(\ref{Mgg}). If we neglect the Higgs boson mass  with 
respect to the messenger masses $m_{L,R}$ in the loop,  we obtain 
\bea
\frac{1}{\Lambda_{\gamma{\subDP}}}\! \!&=&\! 
\frac{\mu_S \sqrt{\alpha \alphaDP} R}{12 \pi} \!\left(\!\frac{
  \sqrt{(m_L^2-m_R^2)^2+4\Delta^2}}{m_L^2m_R^2-\Delta^2}\right)\!\sin{2\theta},
\label{LambdaPDP}
\eea
where $R=N_c \sum_{i=1}^3\left(\eu\qui+\ed\qdi\right)$, 
with $\qui, \qdi$ the $U(1)_{\!D}$ charges in the up and down sectors,
and $\eu=\frac{2}{3}$, $\ed=-\frac{1}{3}$ the corresponding EM charges;  
$\alpha$ is the EM  fine structure constant,
$N_c=3$ is the number of colors, and $\theta$ is the mixing angle diagonalizing Eq.~\rfn{M2}. 
The above result can be easily generalized to 
include the contributions of messengers in the leptonic sector, whose contribution is  $R=e_{\E}\sum_{i=1}^3\left(q_{\EE_i}\right)$, since  in this case $N_c=1$, $\eu=0$, and $\ed=-1$.

A minimal scenario can be realized if we further assume  mass universality in the $S_L$ and $S_R$ messenger sector, with in particular $m_L\simeq m_R\equiv \bar{m}$. Correspondingly, the mixing angle is set to  $\theta=\pi/4$. Then, by defining the mixing parameter $\xi = \Delta/\bar{m}^2$, the eigenvalues of \Eq{M2} become
$
m^2_{\pm}=\bar{m}^2\left(1\pm \xi\right),
$ and the $\Lambda_{\gamma\subDP}$  scale simplifies to 
\bea
\Lambda_{\gamma\subDP} &=& \frac{6\pi v}{R\sqrt{\alpha\alphaDP}}
\frac{1-\xi^2}{\xi^2}\, .
\label{LambdaPDP1}
\eea
In order to avoid tachyons, the mixing parameter should be in the range $0\le \xi \le 1$. However, the upper limit $\xi=1$ is not quite realistic, corresponding to a massless messenger eigevalue. A viable upper limit on $\xi$ can be obtained by requiring that the lightest colored messenger mass satisfies the current lower limit from negative searches of colored scalar fields at the LHC, that we will name $m_B$. In particular, by imposing $m_{-}^2 \!> m_B^2$, one obtains the constraint  $\xi < 1-\frac{m^2_B}{\bar{m}^2}$.
 
One remarkable aspect of the result in Eq.~(\ref{LambdaPDP1}) is the non-decoupling that can show up in the $H\to \gamma \DP$ decay for increasing  messenger masses,  similarly to the $H\to \gamma \gamma$ decay in the SM  in the limit of large top-quark and $W^{\pm}$ masses. In fact, $\Lambda_{\gamma\subDP}$ in Eq.~(\ref{LambdaPDP1}) effectively 
depends only on
one mass scale, \ie \,the Higgs vev (as in the SM two-photon channel), multiplied by a function of two dimensionless free parameters: the mixing parameter $\xi$ and the dark fine structure constant $\alphaDP$. Both parameters can be in principle moderately large (although smaller than 1), regardless of the scale set by the average 
messenger mass $\bar{m}$.
A non-decoupling limit is then realised in the UV regime in which the two mass eigenvalues $m_{\pm}^2$ in the left and right messenger sectors become  arbitrarily large, while keeping fixed (and finite) their relative splitting, expressed by the  $\xi$ parameter. This can indeed occur since the mixing term $\xi =\mu_S v/\bar{m}^2 $ actually depends on two independent mass parameters, the $\mu_S$ scale and the average messenger mass $\bar{m}$.
Hence, keeping $\xi$  finite at large mass scales requires the $\mu_S$ term to scale as $\bar{m}^2/v$ for large  $\bar{m}$. This non-decoupling regime is for instance naturally realized in the model proposed in~\cite{Gabrielli:2013jka} -- on which the simplified dark-sector model assumed here is inspired -- where all the SM Yukawa couplings are radiatively generated by a dark sector. On the other hand, 
as stressed in \cite{Gabrielli:2014oya},  non-decoupling is a general property of the Higgs boson, and does not depend on the peculiar structure of the model in \cite{Gabrielli:2013jka}, provided a messenger sector connecting  the SM and the dark sector exists. 

The messenger interactions can similarly induce new one-loop contributions    to the Higgs decay $H\to \gamma \gamma$, and to the invisible channel $H\to \DP \DP$ arising from decays into two dark photons. 

The corresponding amplitudes have the same structure as  in Eq.~\rfn{Mgg}, and we obtain 
\bea
\Lambda_{\gamma\gamma} = \Lambda_{\gamma\subDP}\, \frac{R}{R_0}
\sqrt{\frac{\alphaDP}{\alpha}}\, , ~~~~
\Lambda_{\subDP\!\subDP} = \Lambda_{\gamma\subDP}
\sqrt{\frac{\alpha}{\alphaDP}}\frac{R}{R_1}\, ,
\eea
where $R_0=3 N_c(\eu^2+\ed^2)$, and $R_1=N_c \sum_{i=1}^3\left(\qui^2+\qdi^2
\right)$, showing analogous non-decoupling properties.

Similar contributions are induced at one loop for the Higgs decay  $H\to Z \DP$,  and for the two-gluon channel $H\to gg$.

When messengers are much heavier than the Higgs boson,  the low-energy Higgs dark-photon interactions can be described by the formalism of effective Lagrangians. By retaining only the relevant low-energy operators,  the corresponding Lagrangian ${\cal L}_{\rm DP_H}$  can then be expressed in terms of (real) dimensionless  coefficients $C_{i\,j}$ (with $i,j =\DP , {\gamma}, Z, g$) as
\bea
 {\mathcal{L}}_{\rm DP_H} &\simeq& \frac{\alpha}{\pi}\Big(\frac{C_{\gamma \subDP}}{v}F^{\mu \nu}{F}^{\scriptscriptstyle{D}}_{\mu\nu} H \, +\,  
\frac{C_{Z\subDP}}{v} Z^{\mu \nu}{F}^{\scriptscriptstyle{D}}_{\mu\nu} H
\, +\, \frac{C_{\subDP\!\subDP}}{v} {{F}^{\scriptscriptstyle{D}}}^{\mu\nu} {F}^{\scriptscriptstyle{D}}_{\mu\nu} H\Big),
\label{Leff}
\eea
where $\alpha$ is the  SM fine structure constant,
and $F_{\mu \nu}$, $Z_{\mu \nu}$, ${F}^{\scriptscriptstyle{D}}_{\mu\nu}$ are the  photon, $Z$-boson, and dark-photon field strengths,  respectively ($F_{\mu \nu}\equiv \partial_{\mu} A_{\nu}-\partial_{\nu} A_{\mu}$ for the photon field, $A_{\mu}$, and analogously for  ${F}^{\scriptscriptstyle{D}}_{\mu\nu}$ and  $Z_{\mu \nu}$).

Additional contributions are induced to the SM Higgs effective interactions
with two photons, one photon and a $Z$, and two gluons, that can be absorbed into the effective Lagrangian ${\cal L}_{\rm SM_H}$ given by
\bea
    {\mathcal{L}}_{\rm SM_H}&\simeq& \frac{\alpha}{\pi}\Big(\frac{C_{\gamma \gamma}}{v}
    F^{\mu \nu}F_{\mu \nu} H \, +\,  
\frac{C_{Z\gamma}}{v} Z^{\mu \nu}F_{\mu \nu} H\Big)
\, +\, \frac{\alpha_S}{\pi}\frac{C_{gg}}{v} G^{a \mu \nu}G^a_{\mu \nu} H,
\label{LeffSM}
\eea
where $\alpha_S$ is the SM strong coupling constant, $G^{a}_{\mu \nu}$ stands for the gluon field strength, and a sum over the color index $a$ is understood. 
Then, for the $C_{i\,j}$ coefficients  one finds
\bea
C_{\gamma\subDP}\! &=&\sqrt{\frac{\alphaDP}{\alpha}}\sum_{i=q,l}\frac{R^i_1}{12}\frac{\xi_i^2}{1-\xi_i^2}\, ,\nonumber\\
C_{\subDP\!\subDP}\!\! &=&\frac{\alphaDP}{\alpha}\sum_{i=q,l}\frac{R^i_2}{12}\frac{\xi_i^2}{1-\xi_i^2}\, ,\nonumber\\
C_{Z\subDP}\!\! &=&\sqrt{\frac{\alphaDP}{\alpha}}\sum_{i=q,l}R^i_{Z\gamma}\frac{R^i_1}{12}\frac{\xi_i^2}{1-\xi_i^2}\, ,\nonumber\\
C_{\gamma\gamma}&=&C^{\rm SM}_{\gamma\gamma}\left(1+\chi \sum_{i=q,l}\frac{R^i_0\xi_i^2}{3F\left(1-\xi_i^2\right)}\right),\nonumber\\
C_{Z\gamma}\! &=&C^{\rm SM}_{Z\gamma}\left(1+ \chi \sum_{i=q,l}R^i_{Z\gamma}\frac{R^i_0\xi_i^2}{3F\left(1-\xi_i^2\right)}\right),\nonumber\\
C_{gg}&=&C^{\rm SM}_{gg}\left(1-\chi\frac{\xi_q^2}{3F_q\left(1-\xi_q^2\right)}\right),
\label{Ci}
\eea
where $C^{\rm SM}_{\gamma\gamma}=\frac{1}{8}F$, 
$C^{\rm SM}_{gg}=   \, \frac{1}{16}F_q\,$,
and the constants $R_{0,1,2}^{q,l}$ are given by
\bea
&& R^q_0=3N_c(e_{\U}^2+e_{\D}^2),~~~~~~~~~~~~~~~~~~~
R^l_0=3\,e^2_{\E}\, , 
\nonumber\\
&&R^q_1=N_c\sum_{i=1}^3\left(e_{\U} q_{\U_i}+e_{\D} q_{\D_i}\right), ~~~~~~~
R^l_1=e_{\E}\sum_{i=1}^3\left(q_{\EE_i}\right)\, , 
\nonumber\\
&&R^{\,q}_2=N_c\sum_{i=1}^3\left(q_{\U_i}^2+q_{\D_i}^2\right), ~~~~~~~~~~~~
R^{\,l}_2\, =\,\sum_{i=1}^3\left(q_{\EE_i}^2+q_{\NN_i}^2\right)\, ,
\label{Ri}
\eea
with $e_{\U}=2/3$, $e_{\D}=-1/3$, and $e_{\E}=-1$,  the  electric charges for up-, down-quarks, and charged leptons, respectively, while  $q_i$ are the corresponding $U(1)_{\!\scriptscriptstyle{D}}$ charges as defined in Table~\ref{tab1}. Here $F,F_F$ and $F_q$ are the usual SM loop factors  given by 
\bea
F=F_W(\beta_W)+F_F\,,~~~~~\;\;\; F_F=\sum_f N_c Q^2_f F_f(\beta_f)\, ,
~~~~~~ F_q=\sum_f F_f(\beta_f)\, ,
\eea
with $N_c=1 (3)$ for leptons (quarks) respectively, 
$\beta_W=4m_W^2/m_H^2$, $\beta_f=4m_f^2/m_H^2$, and
\bea
F_W(x) &=& 2+3x+3x\left(2-x\right)f(x)\, , ~~~~~~
F_f(x) = -2x\left(1+(1-x)f(x)\right)\, ,
\eea
where $f(x)=\arcsin^2[\frac{1}{\sqrt{x}}]$, for $x\ge 1$, and 
$f(x)=-\frac{1}{4}\left(\log\left(\frac{1+\sqrt{1-x}}{1-\sqrt{1-x}}\right)-i\pi\right)^2$,  
for $x< 1$. Including  only the $W^{\pm}$ and top-quark loops in $F$,
we get, for  $m_H=125$ GeV, $F\simeq 6.5$,  $F_t\simeq -1.38$.  The coefficient $\chi=\pm 1$ in Eq.~(\ref{Ci}) parametrizes the relative sign  of the NP and SM contributions in the amplitudes of the $H\to \gamma \gamma$ and $H\to gg$ decays. In our model the $\chi$ sign is a free parameter, since it is related to the relative sign of the SM Higgs vev and the $S$ vev.\footnote{Due to the Bose statistics of the scalar messenger fields, the relative sign of the messenger contributions to $C_{\gamma \gamma}$ (or $C_{Z \gamma}$)  and $C_{gg}$ is anyhow predicted to be negative, as can be checked in Eq.~(\ref{Ci}).}

Concerning the value of  the $R^{q,l}_{Z\gamma}$  constants in Eq.~(\ref{Ci}), this is discussed in more details in \cite{Biswas:2015sha}. In the case of a pair of mass-degenerate down- and up-type colored messengers running in the loop, and in the limit of small mixing, one has $R^{q}_{Z\gamma}\simeq 0.79$, while for a pair of mass-degenerate EW messengers one has $R^{l}_{Z\gamma}\simeq 0.045$.

Notice that, due to the fact that $\xi^2_i \propto v^2$, all the Wilson coefficients in front of the operators in Eqs.({\ref{Leff}) and (\ref{LeffSM}) vanish in the limit of $v\to 0$. This is due to gauge invariance. Indeed, the corresponding SM gauge-invariant effective Lagrangians above the EW scale must require dimension 6 operators, which are obtained by replacing the Higgs field $H$ with $H\to \hat{H}^{\dag} \hat{H}$ in Eqs.({\ref{Leff}) and (\ref{LeffSM}), where $\hat{H}$ is the $SU(2)$ Higgs doublet. Then, after the Higgs field gets the vev, the Lagrangians in Eqs.({\ref{Leff}) and (\ref{LeffSM}) is obtained, with associated Wilson coefficients proportional to $v$.

Finally, by taking into account the parametrization in Eqs.(\ref{Leff}) and (\ref{LeffSM}), one has for the $H\to \gamma \DP$ and $H\to gg$ decay widths~\cite{Gabrielli:2014oya}
\bea
\Gamma(H\to \gamma \DP)&=&\frac{m_H^3 \alpha^2 |C_{\gamma\subDP}|^2}{8 \pi^3 v^2}
\, , ~~~~~
\Gamma(H\to gg)\,=\,\frac{m_H^3 \alpha_S^2 |C_{gg}|^2 (N^2_c-1)}{4 \pi^3 v^2}\, ,
\eea
where $N_c=3$ and $\Gamma(H\to gg)$ is understood to be inclusive in gluons final
states.
Analogous results can be obtained for the  $H\to \DP\DP$,
$H\to Z \DP$,   $H\to \gamma \gamma$ widths replacing  
$|C_{\gamma \subDP}|^2\;$ by $2|C_{\subDP \!\subDP}|^2$,
$|C_{Z \subDP}|^2$, $2|C_{\gamma \gamma}|^2$, respectively.

It is also useful to express the BR's  for   $H\to \gamma \DP,\; \DP \DP, \;  \gamma {\gamma}  $  as a function of the 
{\it relative} exotic NP contribution $r_{i\,j}$ to the  $H\to i \,j$ decay width, as 
the ratio   
\bea
r_{ij}&\equiv&\frac{\Gamma^{\rm m}_{i j}}{\Gamma^{\rm SM}_{\gamma\gamma}}\; ,
\label{rij}
\eea
with $\Gamma^{\rm m}_{i j}$ generically indicating the pure messenger contribution to $H\to  i \,j$, with  $i,j = \gamma , \DP$.
Analogously,  the {\it relative} deviation for the $H\to gg$ decay width will be defined as
\bea
r_{gg}&\equiv&\frac{\Gamma^{\rm m}_{gg}}{\Gamma^{\rm SM}_{gg}}\, .
\label{rgg}
\eea

Then, one obtain the following model-independent parametrization
of the  
$H\to {\gamma}\DP,\; \DP \DP, \;  \gamma {\gamma}  $
BR's as functions of $ r_{ij}$ \cite{Gabrielli:2014oya}
\bea
\BR_{\gamma\subDP}\! &=&\BR^{\rm SM}_{\gamma\gamma}
\frac{r_{\gamma\subDP}}
{1+r_{\subDP\!\subDP}\BR^{\rm SM}_{\gamma\gamma}}\, ,
\nonumber\\
\BR_{\subDP\!\subDP}\!\!\! &=&\BR^{\rm SM}_{\gamma\gamma}
\frac{r_{\subDP\!\subDP}}
{1+r_{\subDP\!\subDP}\BR^{\rm SM}_{\gamma\gamma}}\, ,
\nonumber\\
\BR_{\gamma\gamma}&=&\BR^{\rm SM}_{\gamma\gamma}
\frac{\left(1+\chi \sqrt{r_{\gamma\gamma}}\right)^2}
{1+r_{\subDP\!\subDP}\BR^{\rm SM}_{\gamma\gamma}}\, ,
\label{BRS}
\eea
where, as in Eq.~(\ref{Ci}),  $\chi=\pm 1$  parametrizes the relative sign of the SM and exotic NP amplitudes, and $\BR_{i j} $ stands for  $\BR(H\to i\, j) $. 
As a first approximation, in order to simplify the analysis,
  we have neglected in Eq.(\ref{BRS}) the $r_{gg}$ and $r_{\gamma\gamma}$  contributions to the total width of the Higgs, since they are
  expected to be negligible.

Concerning the Higgs production at the LHC, if colored messenger fields are involved, the cross section from the gluon-gluon fusion modifies as follows
\bea
\sigma_{gg\to H} = \sigma^{SM}_{gg\to H}\left(1-\chi\sqrt{r_{gg}}\right)^2\, .
\label{cross}
\eea
This correction should be taken into account for the colored messengers contribution to the Higgs production from gluon-gluon fusion. In particular,  the  signal strength 
$R_{\gamma \gamma}=\frac{\sigma_{gg \to H}\BR_{\gamma \gamma}}{\sigma^{\rm SM}_{gg \to H}\BR^{\rm SM}_{\gamma \gamma}}$,  will be  given by
\bea
R_{\gamma \gamma}=
\frac{\BR_{\gamma \gamma}}{\BR^{\rm SM}_{\gamma \gamma}}
\left(1-\chi \sqrt{r_{gg}}\right)^2\, .\eea

The model predictions for the ratios $ r_{ij}$ ($i,j =\gamma , \DP$) as defined  in Eq.~(\ref{rij})  
[entering  the  model-independent BR's parametrization in Eq.~(\ref{BRS})],
and $ r_{gg}$ as defined in Eq.~(\ref{rgg})
 are then given by
\bea
r_{\gamma\subDP}&=&2\left(\sum_{i=l,q} X_i R^i_1\right)^2
\left(\frac{\alphaDP}{\alpha}\right)\, ,~~~~~~
r_{\subDP\!\subDP}=\left(\sum_{i=l,q} X_i R_2^i\right)^2
\left(\frac{\alphaDP}{\alpha}\right)^2\, , \\ 
r_{\gamma\gamma}&=&\left(\sum_{i=l,q} X_i R_0^i\right)^2\, , ~~~~~~~~~~~~~~~~~~
r_{gg}=\frac{X_q^2F^2}{F_q^2}\, ,
\label{run}
\eea
where the extra factor 2 in $r_{\gamma\subDP}$ comes from statistics and
\bea
X_{l(q)} \equiv \frac{\xi^2_{l(q)}}{3F(1-\xi_{l(q)}^2)}\, ,  
\label{rdue}
\eea
with $R^{q,l}_{0,1,2}$ defined in Eqs.~(\ref{Ri}).

\begin{figure}[t!]
\begin{center}
\includegraphics[width=3.5in]{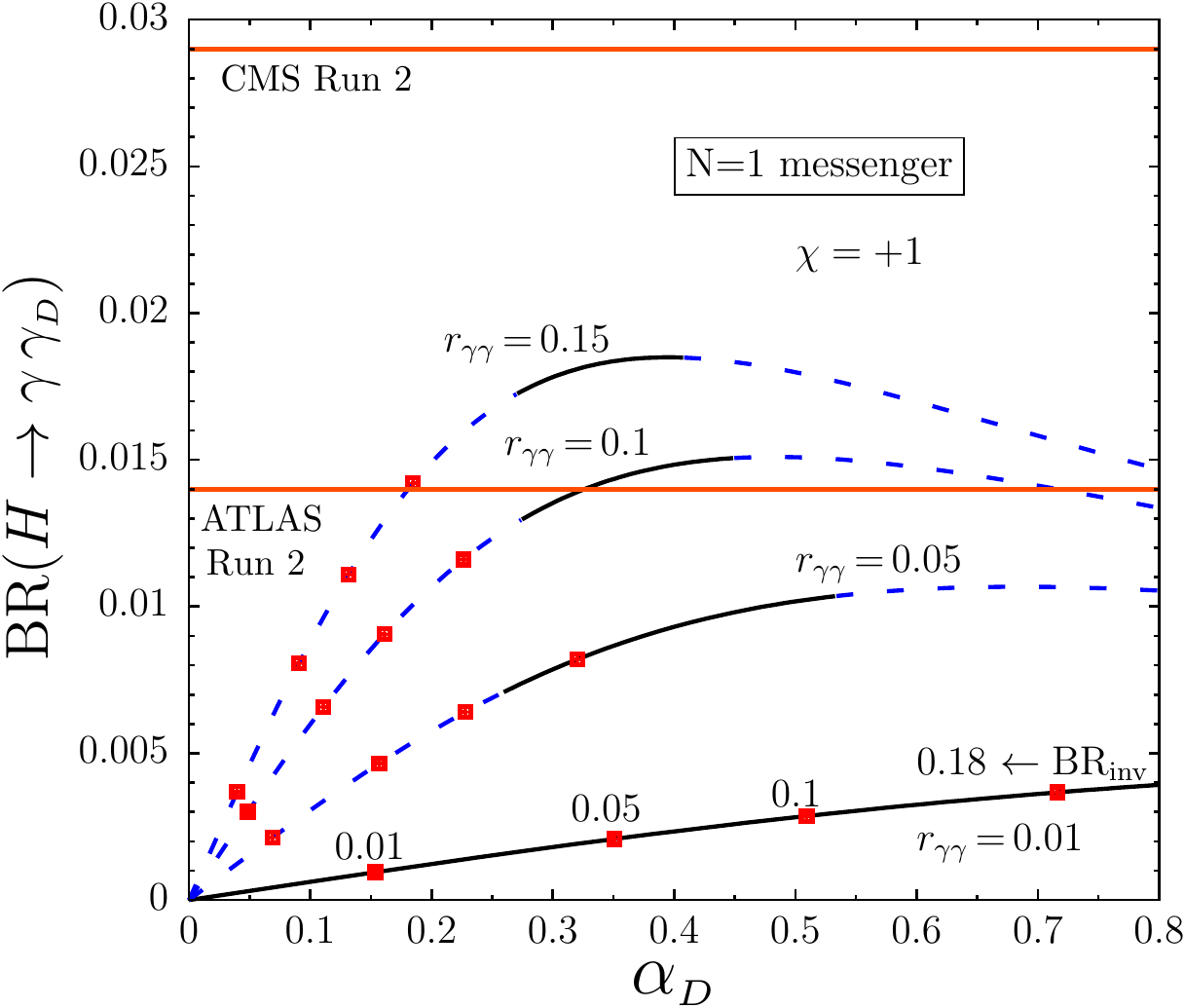} 
\caption{{\small Predictions for $\BR(H\to \gamma \DP)$ versus 
 $\alphaDP$ for different $\BR_{\rm inv}$ and $r_{\gamma\gamma}$,  for the minimal model of 1 (colorless) messenger with unit charges $e=q=1$, and 
 interference coefficient $\chi=+1$. Continuous (dashed) curves are allowed (excluded) by the $\BR(H\to \gamma \gamma)$ measurement at $2\sigma$ level. Horizontal lines indicate the corresponding ATLAS and CMS upper limits on $\BR(H\to \gamma \DP)$ at 95\% CL.
}}
\label{BR}
\end{center}
\end{figure}

Following the analysis in~\cite{Gabrielli:2014oya}, we now consider a minimal model  with only one (colorless) messenger contributing with unit charges $e=q=1$. Updated predictions of this scenario with respect to \cite{Gabrielli:2014oya} are reported in  Fig.~\ref{BR}, where  we plot $\BR(H\to \gamma \DP)$ versus $\alphaDP$. The curves are evaluated for 
$r_{\gamma\gamma}=0.01,\, 0.05\, , 0.1\, , 0.15$, corresponding to mixing parameter $\xi=0.81,0.90,0.92,0.94$ respectively. 
The red squares correspond to  different  $\BR_{\subDP\subDP}$ values (increasing from left to right), with the $H\to \DP\DP$ decay assumed to provide the leading contribution to the Higgs invisible branching ratio BR$_{\rm inv}$. The value BR$_{\rm inv}=0.18$ correspond to the current experimental upper bound at 95\% C.L. from CMS \cite{CMS:2022qva}, which is less stringent than the corresponding one from ATLAS \cite{ATLAS:2022yvh} (BR$_{\rm inv}<0.14$). Then, the points to the right of the red square with BR$_{\rm inv}=0.18$ can be assumed (conservatively) to be excluded at 95\% C.L. from the current limits on  BR$_{\rm inv}$.

The full lines in Fig.~\ref{BR} correspond to the allowed values of $\BR_{\gamma\gamma}$ from the current limits on signal strengths at $2\sigma$ level~\cite{ParticleDataGroup:2020ssz}
\bea
0.93\;\le R_{\gamma\gamma}\le 1.31\, ,
\label{Rgaga}
\eea
while the dashed lines correspond to predictions outside that range. 
For the SM central value we used
$\BR^{\rm SM}_{\gamma\gamma}=2.27\times 10^{-3}$\cite{ParticleDataGroup:2020ssz}. The horizontal (orange) bands are  the observed upper limit on $\BR_{\gamma \subDP}$ at 95\% C.L. from the ATLAS (1.4\%) \cite{ATLAS:2021pdg} and CMS (2.9\%) \cite{CMS:2020krr} analyses (these limits will be  discussed in more details in section 4.2).
We assume constructive interference between exotic and  SM contributions (\ie,   $\chi=1$). 
Due to the asymmetry of the range in Eq.~(\ref{Rgaga}) with respect to the $R_{\gamma\gamma}$ SM value,
the experimental $\BR_{\gamma\gamma}$ constraints are correspondingly less effective, thus allowing a wider $\BR_{\gamma\subDP}$ range.

In Fig.~\ref{BRN6mess} we show the corresponding results, for a non-minimal model consisting of N=6 EW messengers [$SU(3)_c$ color singlet] (left plot), and a $SU(3)_c$ color triplet (right plot), with SM QN as in Table~\ref{tab1}, and universal unitary $U(1)_{\!\scriptscriptstyle{D}}$ charges 
 ($q_{E_i}=q_{N_i}=q_{D_i}=q_{U_i}=1$) for all messengers.
The same notations  as in Fig.~\ref{BR}  for the curves and red-square points are adopted. Constructive interferences between exotic and SM contributions are assumed ($\chi=1$). Curves are shown for  $r_{\gamma\gamma}=0.005,0.02,0.05,0.1$, corresponding to  universal mixing parameters 
$\xi^{l}=0.56,0.69,0.76,0.82$, and  $\xi^{q}=0.47,0.60,0.68,0.74$, in the left and right plot, respectively. Note that, 
in Fig.~\ref{BRN6mess} (right plot), the $\BR_{\gamma\gamma}$ constraints take into account the messenger contribution to the gluon-gluon Higgs production cross section in the signal strength~$R_{\gamma \gamma}$.

\begin{figure}[t!]
  \begin{center}
  \includegraphics[width=3in]{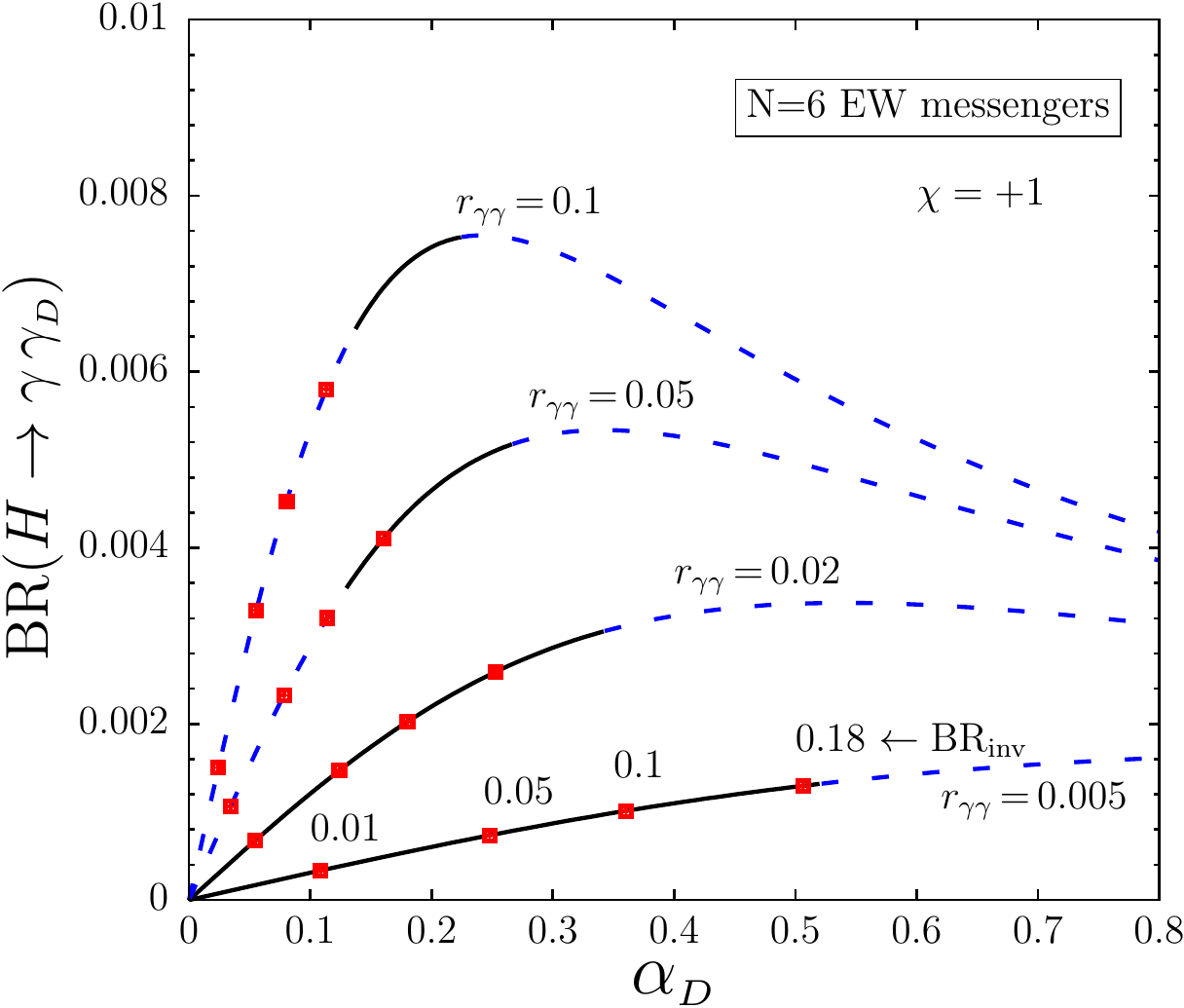}
  \includegraphics[width=3in]{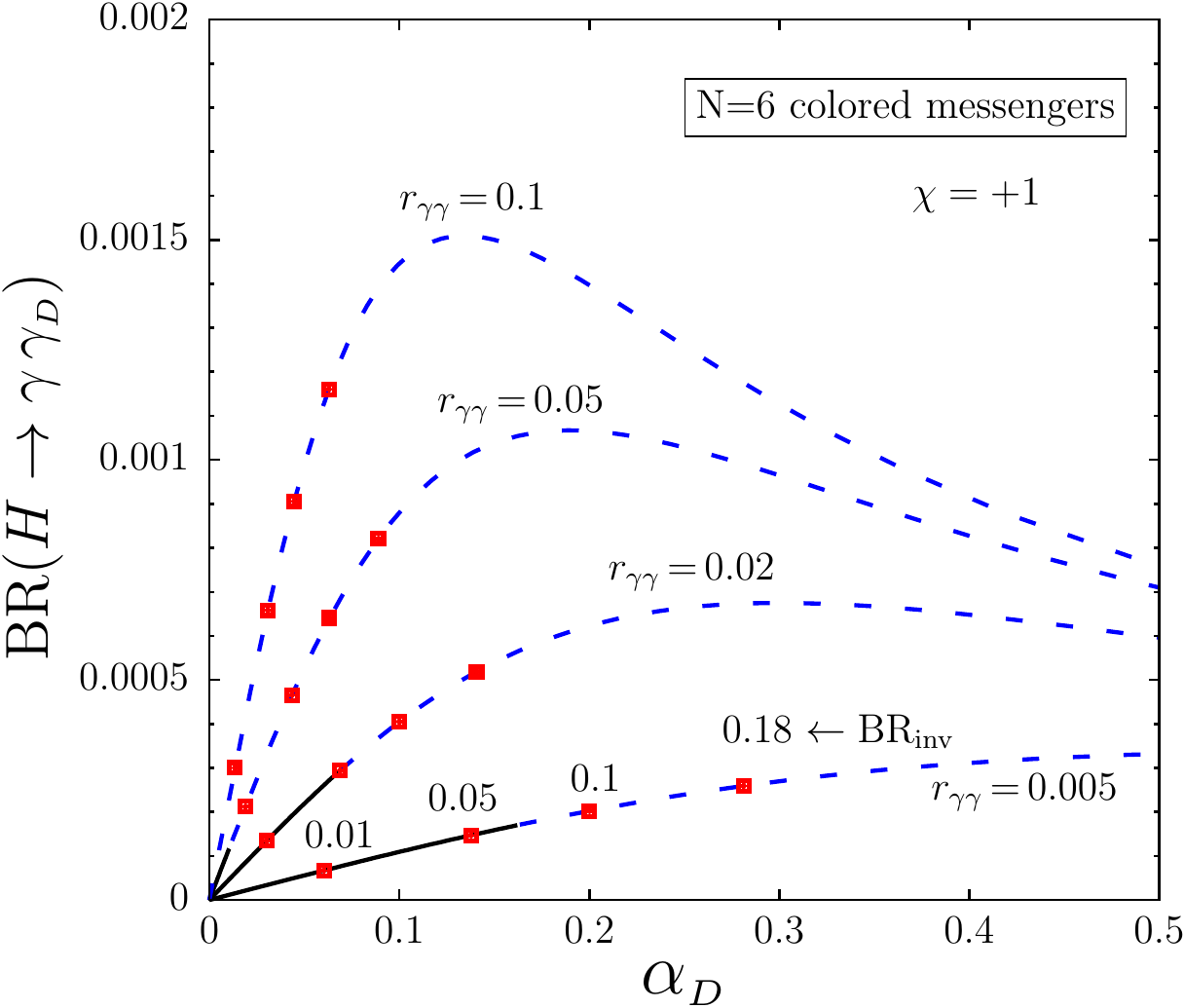}
  \caption{{\small Predictions for $\BR(H\to \gamma \DP)$ versus $\alphaDP$, as in Fig.{\ref{BR}}, for extended portal models with N=6 EW $SU(3)_c$ singlet messengers (left plot) and N=6 $SU(3)_c$ triplet messengers (right plot), with same SM QN of squarks and sleptons, and interference coefficient $\chi=+1$.
}}
\label{BRN6mess}
\end{center}
\end{figure}
As we can see from these results, the allowed $\BR(H\to \gamma \DP)$ for the minimal model is below~$1\%$, consistently with all model parameters and current LHC constraints. On the other hand, the allowed  $\BR(H\to \gamma \DP)$  is reduced to less than $4\times 10^{-3}$ and $3\times 10^{-4}$ for the case of N=6 EW and colored messengers, respectively. Indeed, increasing the number of messengers  at fixed $r_{\gamma \gamma}$, $\BR(H\to \gamma \DP)$ decreases, since  the larger the number of messengers the larger the contribution to the invisible rate given by $H\to \DP \DP$ in Eq.~(\ref{BRS}), thus raising the total width and lowering  $\BR(H\to \gamma \DP)$.

A major result of this analysis is that the current sensitivity in the  $\BR(H\to \gamma \DP)$ measurement by ATLAS and CMS   is presently almost one order of magnitude weaker than what is needed for detecting $\BR(H\to \gamma \DP)$ in the allowed range, which is consistent with actual constraints on $\BR(H\to \gamma \gamma)$ and $\BR(H\to {\rm invisible})$. 
The present SM agreement of the latter measurements  indicates that more Higgs data are needed in order to explore the allowed  $\BR(H\to \gamma \DP)$ range at a few permil level.

\subsection{About the spin of the invisible dark photon}
We now investigate whether the observation of the monochromatic photon signature discussed above could uniquely identify the dark-photon production. Because of the isotropic nature of a scalar  decay, 
in the $H\to\gamma X$ channel  
it is not possible to disentangle the spin nature of a dark $X$ {\it boson}, if $X$ is detected as missing energy (note that a {\it fermionic} $X$ particle would violate Lorentz invariance). Indeed, in the latter case, one cannot reconstruct $X$ spin properties via kinematics of its visible decay products as in visible decays. 
Actually, we will see that that identification of such a signature with a dark {\it photon} (hence with a spin=1 field) is the most realistic.
In particular, we will discuss below possible scenarios of NP that could fake the dark-photon signature,  estimate their corresponding BR, and find that the dark-photon  $ H\to \gamma \DP$ interpretation of the $H\to\gamma X$ decay is by far the {\it most viable}.

Let us start with the possibility that the $X$ particle is either a scalar or a pseudo-scalar particle (for instance, an axion-like particle). According to the angular momentum conservation, the Higgs boson cannot decay into a photon plus a scalar or pseudo-scalar particle, ruling out the possibility that $X$ is a scalar or an axion-like particle. Indeed, by considering the two-body $H\to \gamma X$ decay in the rest frame of the Higgs boson, one can  see that the (zero) helicity of the initial state cannot be conserved in the final state, due to the photon $h=\pm 1$ helicity, for scalar/pseudoscalar $X$'s. This is also manifest in the effective Lagrangian approach, when  trying to build a gauge invariant $H S\gamma$ interaction ($S$ standing for a generic  scalar or pseudoscalar field). Indeed, this kind of interaction always vanishes for on-shell fields, up to a total derivative. In particular, the Lagrangian is proportional to the following Lorentz and gauge invariant term $(\partial_{\mu} H)(\partial_{\nu}S) F^{\mu\nu}$, which is equivalent (up to a total derivative) to the sum of the $(\partial_{\mu} \partial_{\nu} H)S F^{\mu\nu}$ and $(\partial_{\mu}H) S \partial_{\nu} F^{\mu\nu}$ terms. The first term vanishes for the antisymmetric property of the  $F^{\mu\nu}$ tensor under the $(\mu,\nu)$ indices exchange, while the second term vanishes for on-shell photon fields due to the condition $\partial_{\nu}F^{\mu\nu}=0$. Analogous conclusions hold for other terms with different combination of derivatives.

As next potential candidate for the $X$ boson in the $H\to\gamma X$ decay, we consider a massive spin-2 field $X=G$ which is universally coupled to the total energy-momentum $T_{\mu\nu}$ of SM fields and of any potential NP beyond it. This is characterized by a rank-2 symmetric and traceless tensor field $G_{\mu\nu}$ associated to the spin-2 particle.  As in the case of a massive graviton, this coupling reads
\bea
L_{\G}=-\frac{1}{\Lambda_{\G}} T^{\mu\nu}G_{\mu\nu}\, .
\label{Lspin2}
\eea
Since we assume $G_{\mu\nu}$ not to be related to gravitational interactions, we take  the effective scale $\Lambda_{\G}$ as a free parameter, uncorrelated with the Planck mass, and of the order of the ${\rm TeV}$ scale. This scale turns to the well known $\Lambda_{\G}^{-1}=\sqrt{8\pi G_N}$ in the ordinary case of a massless graviton in the General Relativity, with $G_N$ the Newton 
constant~\footnote{We do not make any hypothesis on the origin of such spin-2 field, limiting ourselves to the linear theory in flat space, avoiding to enter into the issue of a consistent theory of massive spin-2 fields related to the non-linear massive graviton interactions, since these do not affect the results presented here.}. 
The free Lagrangian for the massive spin-2 is then given by the Fierz-Pauli Lagrangian~\cite{Fierz:1939ix}. The corresponding Feynman rules for the $G$ interaction in Eq.~(\ref{Lspin2}) can be derived, for instance, from literature on quantum gravity models in large extra-dimensions, where massive Kaluza-Klein graviton fields appear \cite{Han:1998sg},\cite{Giudice:1998ck}.

The coupling in Eq.~(\ref{Lspin2}) is sufficient to generate new finite  contributions at loop level for the effective $HG\gamma$ coupling entering  the $H\to \gamma \,G$ decay. Indeed, due to the fact that  $G_{\mu\nu}$ is coupled to the conserved  energy-momentum tensor $T^{\mu\nu}$ of matter fields, the theory is renormalizable against radiative corrections of SM matter fields only, provided   $G_{\mu\nu}$ is taken as an external 
on-shell field.

From basic kinematical considerations,  $H\to \gamma \,G$ is now allowed by  angular momentum conservation, since a massive spin-2 particle has 5 spin components, corresponding to $S_z=-2,-1,0,1,2$ (with $S_z$ standing for the usual eigenvalues of the spin component along the $z$-axis). However, only the $h=\pm 1$ helicity states of the massive spin-2 components will contribute to the decay. On the other hand, for a massless spin-2 field (like the Einstein graviton) the reaction is forbidden since the graviton has only two helicity states $h=\pm 2$, and the corresponding  decay amplitude will vanish.
Since the massless limit for the amplitude should be recovered from the massive spin-2 case for vanishing masses,  the rate of the  $H\to \gamma \, G$ is expected to be suppressed by terms of the order of $m_G^2/m_H^2$.

In order to check these expectations, we provide below the most general Lorentz and gauge invariant structure of the  $M(H\to \gamma \,G)$ amplitude for the decay
\bea
H\to \gamma(k)\, G(q)\, 
\eea
that, to our knowledge, is not yet present in the literature, and can  be expressed as 
\bea
M(H\to \gamma \, G)&=&\hat{M}_{\mu\alpha\beta}(p,q) \,\epsilon^{\mu}(k)~
\epsilon^{\alpha\beta}(q)\, .
\label{MGH1}
\eea
Here, $\epsilon^{\mu}(k)$ and $\epsilon^{\alpha\beta}(q)$ are the corresponding polarization vectors for the on-shell photon and massive graviton $G$, respectively, with $\epsilon^{\alpha\beta}(q)$ a symmetric and traceless spin-2 tensor, satisfying the on-shell conditions $g_{\alpha \beta} \epsilon^{\alpha\beta}(q)=q_{\alpha}\epsilon^{\alpha\beta}(q)=q_{\beta}\epsilon^{\alpha\beta}(q)=0$, with $g_{\mu\nu}$ the Minkowski metric. Then $\hat{M}_{\mu\alpha\beta}(p,q)$ can be parametrized as follows
\bea
\!\!\!\!\!\hat{M}_{\mu\alpha\beta}(p,q) &=&
F_G \left[ q_{\mu}\left(k_{\alpha}q_{\beta}+k_{\beta}q_{\alpha}\right)
-2\left(\frac{m_G^2} {k\!\cdot\! q}\right)q_{\mu}k_{\alpha}k_{\beta}
+\left(m_G^2-k\!\cdot\! q\right)
\left(g_{\mu\alpha}k_{\beta}+g_{\mu\beta}k_{\alpha}\right)
\right]\!,~
\label{MGH2}
\eea
where $F_G$ is a form factor  having ${\rm [mass]}^{-2}$ dimension (which absorbs also the electromagnetic couplings), depending only on the Higgs mass and $m_G$. The $F_G$ form factor, which is expected to arise at loop level  from the interaction in Eq.~(\ref{Lspin2}) (see below), is free from power $m_G\to 0$  infrared singularities of the type $1/m_G^2$, since no $G$ field is propagating in the loop.

It is easy to see that  $\hat{M}_{\mu\alpha\beta}(p,q)$ in Eq.~(\ref{MGH2}) satisfies the following Ward Identities (WI)
\bea
k^{\mu} \hat{M}_{\mu\alpha\beta}(p,q) &=& q^{\alpha} \hat{M}_{\mu\alpha\beta}(p,q)\,=\,q^{\beta} \hat{M}_{\mu\alpha\beta}(p,q)\,=\,0\, ,
\label{WI}
\eea
including the (traceless) additional condition $g^{\alpha\beta}\hat{M}_{\mu\alpha\beta}(p,q)=2F_Gk_{\mu}$, that vanishes when contracted with $\epsilon^{\mu}(k)$ for on-shell photons. The above WI are a consequence of the gauge invariance of the amplitude in Eq.~(\ref{MGH1}) under  gauge transformations of the theory, that in the momentum space read :  $\epsilon^{\mu}(k)\to \epsilon^{\mu}(k) + k^{\mu}$, $\epsilon^{\alpha\beta}(q)\to \epsilon^{\alpha\beta}(q)+q^{\alpha}\varepsilon^{\beta}+q^{\beta}\varepsilon^{\alpha}-1/2g^{\alpha\beta}q\!\cdot\!\varepsilon$ (with  $\varepsilon$ a generic 4-vector).

Finally, by summing over photon and spin-2 polarizations and integrating over the final phase space (see \cite{Han:1998sg,Giudice:1998ck} for the expression of the polarization matrix of a massive spin-2  field), the total width for the $H\to \gamma G$ decay is  given by
\bea
\Gamma(H\to \gamma G)=\frac{F_G^2 m_H^3 m_G^2}{16 \pi}\left(1-r_G\right)^3 ,
\eea
where $r_G=m_G^2/m_H^2$. As we can see from these results, the above width vanishes in the $m_G\to 0$ limit, as expected from angular momentum conservation. 

We stress that the amplitude in Eqs.~(\ref{MGH1},\ref{MGH2}) cannot arise at  tree level, and  is expected to be induced only by higher-order contributions in perturbation theory.
In particular, since $H\to \gamma G$ is a $C$-parity violating process, one can easily check that, due to the $C$-parity conservation of electromagnetic interactions, its contribution exactly vanishes at  one loop in the SM and beyond. Then, a (finite) non-vanishing contribution to the $F_G$ form factor  can only arise starting from the next-to-leading order at two loops, due to potential corrections induced by $C$-parity violating interactions\footnote{Notice that, thanks to the WI in Eqs.(\ref{WI}), the corresponding UV contribution is finite at any order in perturbation theory within the SM and in any of its NP extensions, provided the spin-2 field acts only as an external classical source without propagating in the loops.}. The computation of this effect at two loops in the SM goes anyhow beyond the purpose of the present review.

We will now show that  BR$(H\to \gamma G)$  is in general expected to be too small to be observable.
From dimensional grounds one can see that the loop induced $F_G$
form factor  should be proportional to $\sim \alpha/\Lambda_{\G}^2$ with $\Lambda_{\G}$ defined in Eq.~(\ref{Lspin2}) (neglecting  both the loop suppression factors at denominator and other coupling products) which implies that the total width $\Gamma(H\to \gamma \,G)$ is proportional to $ \sim \alpha\, m_H^3 m_G^2/\Lambda_{\G}^4$ .
As shown in\cite{Comelato:2020cyk}, the $\Lambda_{\G}$ effective scale is expected to be not smaller than  $(1-100)$ TeV 
(depending on the value of the graviton mass) for light invisible spin-2 fields with masses between  the eV and the  GeV scale,  and  even heavier  for larger masses  (for more details see~\cite{Comelato:2020cyk}). For  $m_G$ $\lsim 100$ MeV, the corresponding BR would be too small to be observed even for $\Lambda_{\G}\sim 1\, {\rm TeV}$,  hence strongly disfavouring any massive spin-2 explanation for the $H\to \gamma X$ signal.

Finally, the above arguments could be extended --{\it cum granus salis}-- to show that also BR($H\to \gamma X_S$), with $X_S$ a dark boson with spin $S>2$, is expected to be strongly suppressed.
Although, there is not any consistent S-matrix theory for interacting higher spin fields with $S>2$, we can estimate the corresponding BR
using angular momentum conservation. The argument is the following. For massless $X_S$ particles with spin $S>1$ in $D=4$ dimensions, only the two $h=\pm S$ helicity states are available\footnote{This result follows from the number of helicity states $n_h$ for a massless particle of spin $S$ in $D$ dimensions, given by $n_h=(D+2S-4) (S+D-5)! / ( S! (D-4)! )$, that for $D=4$ is always $n_h=2$~\cite{anselmi}.}.
Then, as for the massless spin-2 case discussed above, the Higgs boson cannot decay into a photon plus a massless $X_{S > 1}$ boson due to angular momentum conservation. Therefore, it is expected that also in this case, for massive higher spin particles, the decay can only proceed via its  $S_z=\pm 1$ spin components. However, the corresponding $S_z=\pm 1$ contributions to the amplitude should vanish in the $m_G\to 0$ limit in order to reproduce the massless limit. Therefore, also for  $S>2$, we expect the width to be strongly suppressed by terms of order $m_S^2/m_H^2$, thus recovering the same conclusions as for a light spin-2 $X$ boson state.

Apart from the  two-body decays just discussed, there is  the possibility that the two-body signature might be faked by three-body final states with one photon plus missing energy. In particular, three-body final states 
with  two invisible particles, one of which very soft, can show up with  an almost resonant monochromatic photon, with energy $\sim m_H/2$ in the Higgs rest frame, plus missing energy. This case has been considered for instance  in~\cite{Beauchesne:2022svl} in the framework of SUSY models. In this context, the final state is generated in two steps. First the Higgs boson decays into a neutralino ($N$) plus a light (invisible) gravitino ($\tilde{g}$), $H\to N \tilde{g}$. Then the neutralino decays into a photon plus gravitino, $N\to \tilde{g} \gamma$, with the two gravitinos giving missing energy in the detector.
This signature can fake the dark-photon one only if the neutralino is not much lighter then the Higgs boson, so that one of the gravitinos is very soft and goes  undetected. However, as shown in~\cite{Beauchesne:2022svl}, the LHC can almost rule out this possibility at the 95\% CL, depending on the integrated luminosity and branching ratios of SUSY decays.

In conclusion, a monochromatic photon signature in the Higgs  
$H\to \gamma X$ decay would 
in practice uniquely identify the $X$ particle as a dark photon.

\section{Dark photon production in Higgs decays at LHC}

In this section we summarize the main results of our phenomenological studies \cite{Gabrielli:2014oya, Biswas:2016jsh} of the dark-photon production via Higgs-decay at the Large Hadron Collider experiment at CERN.
LHC is the world's largest particle accelerator till date where {\it proton-proton} collisions take place at high center-of-mass (c.m.) energies. The main Higgs production channels in proton-proton collisions are the Higgs production
via gluon fusion (ggF), VBF and associated production (VH).

\begin{figure}[t!]
  \begin{center}
  \includegraphics[width=2.6in]{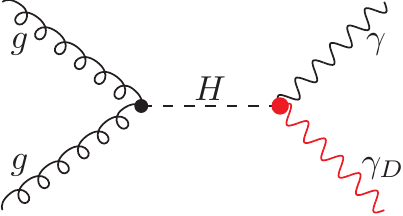}
\hskip 30pt  \includegraphics[width=2.0in]{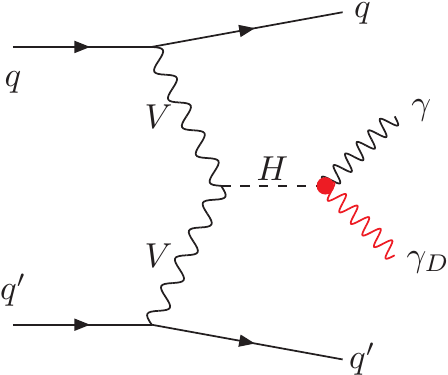}
  \caption{{\small Feynman diagrams for the Higgs production via  gluon fusion ({\it left}), and  VBF ({\it right}) channels at the LHC.
}}
\label{fig:FD_Higgs-LHC}
\end{center}
\end{figure}

\subsection{Gluon fusion production}
Higgs production via gluon fusion process is one of the dominant modes at the LHC. The estimated cross section in this channel is 49.85 (19.37) pb at 14 (8) TeV c.m. energy, and gives the largest production rate for a hypothetical scalar boson 
with SM Higgs like couplings for the entire mass range of interest \cite{LHCHiggsCrossSectionWorkingGroup:2013rie}. We have simulated the $pp\rightarrow H\rightarrow \gamma\DP$ process, where the Higgs is produced in the gluon-fusion 
channel both at $8$ TeV and $14$ TeV c.m. energies. The Feynman diagram for this process is depicted in Fig.~\ref{fig:FD_Higgs-LHC} (left).
The signal is characterized by {\it a single photon recoiling against missing transverse momentum} ($\gamma + \slashed{E}_T$). The SM backgrounds for this process 
are dominated by $pp\rightarrow \!\gamma j$ and QCD multi-jet background  $p p \rightarrow {\rm ~jets}$, where the missing transverse momentum can arise from a number of sources, {\it e.g.}, a) jet energy mismeasurement, 
b) invisible neutrinos arising from decays of heavy-flavor jets, and c) very forward particles escaping the detector.
The latter process contributes to the $\gamma + \slashed{E}_T$ final state whenever one of the jets is misidentified as a photon. The main electroweak background consists of the channels $pp \rightarrow W \rightarrow e\nu$, where the electron is misidentified as a photon, $pp\rightarrow W(\rightarrow \ell\nu) \gamma$, for $\ell$ outside charged-lepton acceptance, and $pp\rightarrow Z(\rightarrow \nu\nu) \gamma$.

We have simulated both the parton level signal and background events in the
context of gluon fusion process using ALPGEN (v2.14) event generator \cite{Mangano:2002ea}. The
signal processes generated by ALPGEN consists of $pp \to H$ and $pp \to Hj$ whereas those for the backgrounds are $pp \to \gamma j$ and $pp \to jj$. The other electroweak backgrounds such as $pp \to W$ , $pp \to W \gamma$ and
$pp \to Z\gamma$ are generated using Mad-Graph5 aMC@NLO (v2.2.2) \cite{Alwall:2014hca}. These events are then interfaced with PYTHIA (v6.4) \cite{Sjostrand:2006za} for parton shower, hadronization and clustering of these hadrons to get jets using simple cone algorithm. More importantly, the decay of the Higgs in to a photon and dark photon has also been ensured at the PYTHIA level using appropriate branching fraction. We have also implemented finite detector resolution effect on the final state reconstructed objects assuming a Gaussian smearing function.

Several kinematic observables, such as missing transverse energy ($\slashed{E}_T$),  transverse momentum of the photon ($p_T^{\gamma} $), transverse mass of the photon-invisible system ($M^{T}_{\gamma\subDP}$) have been proposed to isolate
the signal from the SM backgrounds ~\cite{Gabrielli:2014oya, Biswas:2016jsh}. The transverse-mass variable that carries the typical signature of the $H\!\to \!\gamma\DP$ decay is defined as $M^{T}_{\gamma\subDP}=\sqrt{2p_T^\gamma \slashed{E}_T(1-\cos\Delta\phi)}$, where   $\Delta\phi$ is the azimuthal distance between the photon transverse momentum $p_T^\gamma$, and the missing transverse momentum $\slashed{E}_T$. The $\slashed{E}_T$ is defined as the unbalanced momentum 
in the transverse plane due to the presence of the invisible particles. 
Fig.~\ref{fig:transv_mass} (left plot) depicts the expected distribution of the $M^{T}_{\gamma\subDP}$ variable for the signal and SM backgrounds in the gluon fusion channel.

\begin{figure}
\centering
\includegraphics[width=0.47\textwidth]{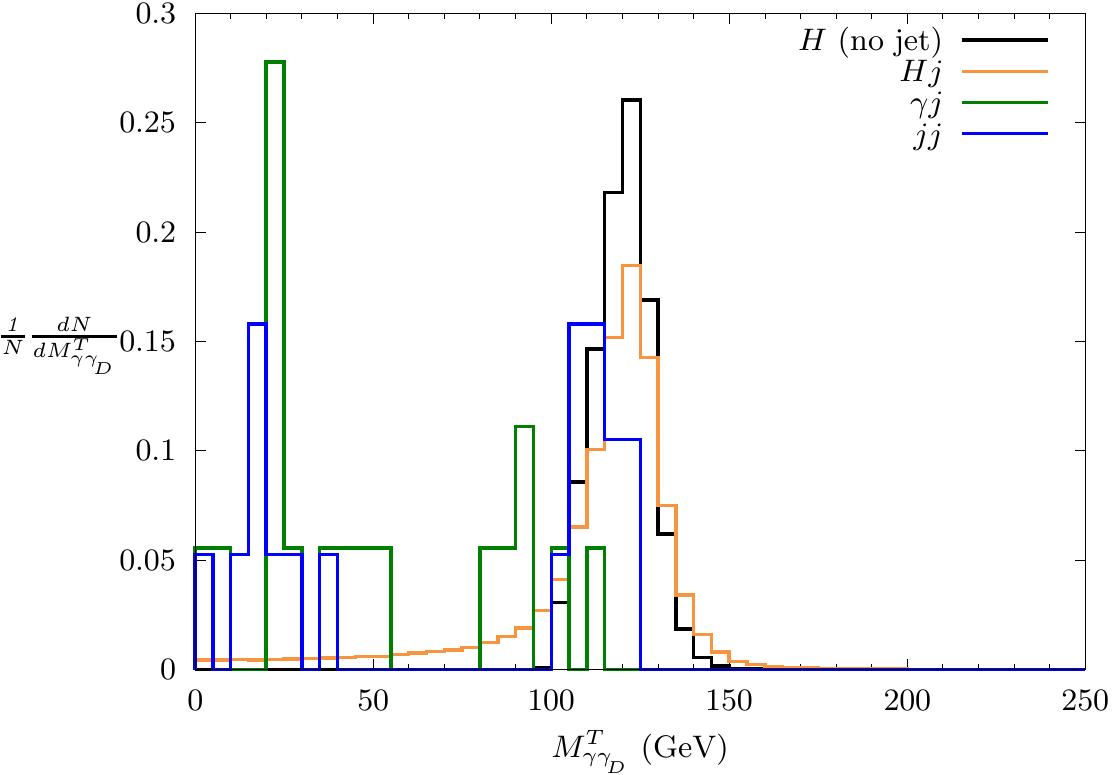}
\includegraphics[width=0.47\textwidth]{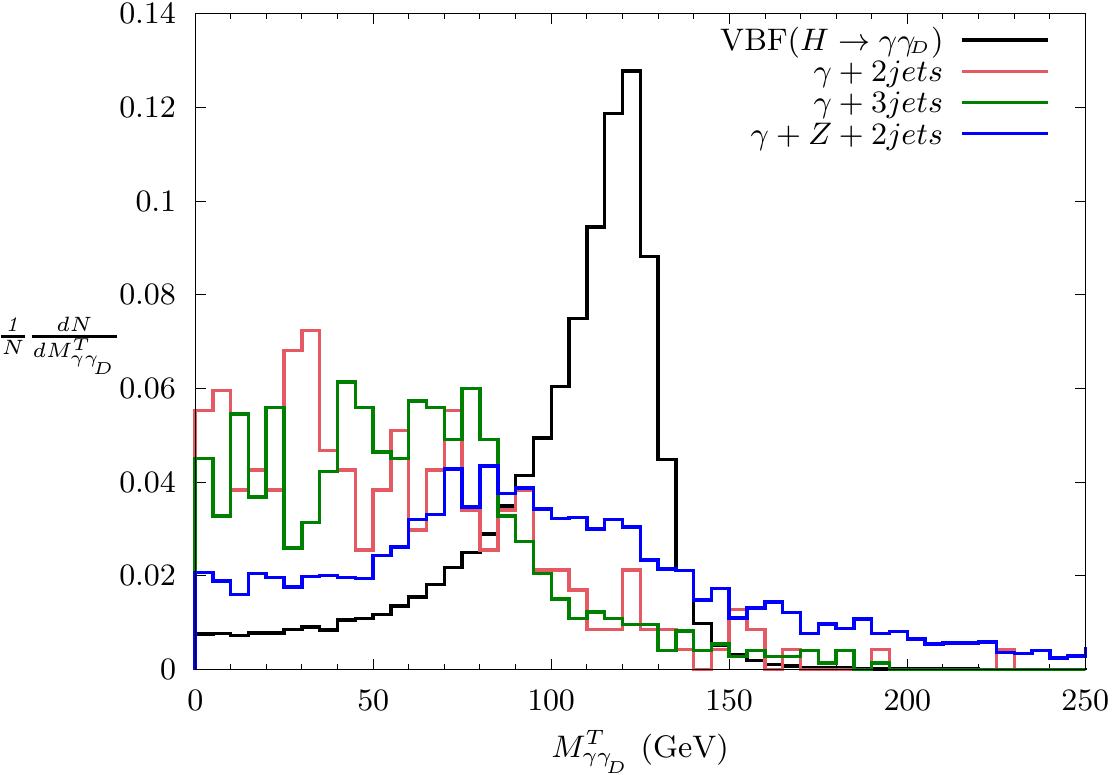}
\caption{Transverse-mass distributions for the $H\!\to \!\gamma\DP$ signal in the gluon-fusion process ({\it left}), and  VBF process ({\it right})~\cite{Biswas:2016jsh}. Corresponding distributions for SM backgrounds for  
inclusive $\gamma+\slashed{E}_T$ (ggF) and $\gamma+\slashed{E}_T+$ two forward jets (VBF) final states (with no isolated leptons), respectively, have also  been shown. All distributions are normalized to unity.}
\label{fig:transv_mass}
\end{figure}

The results of our simulation is summarized in Table ~\ref{tab:ggf} in terms of the cross section times cut efficiency ($\sigma \times A$) of the signal and the background processes after the implementation of the event selection criteria detailed in~\cite{Biswas:2016jsh} 
at two different c.m. energies, 8~TeV and 14~TeV, assuming BR($H\!\to \!\gamma\DP$) $\sim 1\%$. Due to large QCD backgrounds and poor estimate of the missing transverse momentum arising from jet energy mismeasurement one has less control over the SM background in this channel.
Using as significance estimator $S/\sqrt{S+B}$, the  analysis at 8~TeV with the 20 fb$^{-1}$ data set gives a $5\sigma$ discovery reach at BR$_{\gamma\subDP}\simeq 4.8\times 10^{-3}$.

In Table \ref{tab:signf}, we present the $2\sigma$ exclusion limit and the $5\sigma$ discovery potential for BR$_{\gamma \subDP}$  after extrapolating the optimization technique to 14 TeV center of mass energy at three different integrated luminosities. At an integrated luminosity of
100 (300)fb$^{-1}$ the $2\sigma$ exclusion limit on BR$_{\gamma \subDP}$ found to be
$6.4 \times 10^{-4}  (3.7 \times 10^{-4}$), whereas the $5\sigma$ discovery reach is
$1.6 \times 10^{-3} (9.2 \times 10^{-4})$. The corresponding $5\sigma$ 
reach can be improved down to $2.9 \times 10^{-4}$  at the High-Luminosity LHC (HL-LHC), with an integrated luminosity of 3 ab$^{-1}$~\cite{Biswas:2016jsh}."

\begin{table}
\begin{center}
\begin{tabular}{c|c|c}
 & $\sigma\times A$ \small{ [8 \!TeV]} & $\sigma\times A$\small{ [14 \!TeV]} \\ \hline
$H\!\to\! \gamma\DP\;\;$\small{ (BR$_{\gamma\subDP}=1\%)$} & 44 & 101 \\ \hline
 $\gamma j$ & 63 & 202 \\
 $jj\rightarrow\gamma j$ & 59 & 432 \\
 $e\rightarrow \gamma$ & 55 & 93 \\
 $W(\rightarrow \!\ell\nu) \gamma$ & 58 & 123 \\
 $Z(\rightarrow \!\nu\nu) \gamma$ & 102 & 174 \\ \hline
\small{total background} & 337 & 1024 \\ \hline
\end{tabular}
\caption{Cross section times acceptance $A$ (in fb) for the gluon-fusion signal  and corresponding SM backgrounds at 8 and 14 TeV, assuming BR$_{\gamma \subDP}\!\!\!=\,$1\%, with the selection \mbox{$p_T^\gamma > 50\ {\rm GeV}$, $|\eta^\gamma|<1.44$}, \mbox{$\slashed{E}_T > 50$ GeV}, and \mbox{$100\ {\rm GeV} < M^{T}_{\gamma\subDP} < 130\ {\rm GeV}$}~\cite{Biswas:2016jsh}.}
\label{tab:ggf}
\end{center}
\end{table}

\subsection{Vector boson fusion (VBF) production}

Here we discuss the phenomenological study of the Higgs production in
the VBF process and its subsequent decay to $H\rightarrow\gamma\DP$. 
The VBF production channel is the most dominant mode of Higgs production at the LHC after ggF
with an estimated cross section of 4.18 (1.578) pb at 14 (8) TeV c.m. energy \cite{LHCHiggsCrossSectionWorkingGroup:2013rie}.
The final state in this case ($pp \rightarrow Hjj\rightarrow\gamma\DP jj$) 
consists of an isolated photon, missing transverse energy and two forward jets with opposite rapidity. 
Corresponding Feynman diagram is shown in Fig.~\ref{fig:FD_Higgs-LHC} ({\it right}).
The SM background contributions to this final state mainly come
from $\gamma+$jets, QCD multi-jets, and $\gamma+Z(\to\!\bar\nu \nu)+$jets processes.
Given the magnitude of QCD multi-jets background even a jet faking as photon with a mistagging rate of~0.1\% gives dominant background contribution to our final state
as discussed in the $ggF$ study. The identification efficiency of a true photon assumed to be same as that is used in the $ggF$ process.

In this case, the parton level signal events for $pp \to Hjj$ via
VBF channel are generated using Madgraph event generator
whereas the parton level background events for the processes
$pp \to \gamma + jets$, QCD multijets and
$pp \to \gamma + Z + jets$ are
all generated using ALPGEN event generator. The remaining steps such as
parton showering, hadronization, decay and clustering of hadrons in to
jets are all performed using PYTHIA.
The energy-momenta of the final state reconstructed objects have been smeared in a similar manner.

The results of our phenomenological analysis is presented in Table \ref{vbf1} which contains the cross sections times cut acceptance for the signal and dominant SM backgrounds after 
the sequential application of basic cuts, rapidity cuts on the two forward jets, and transverse-mass cut on the photon plus missing transverse-energy system as described earlier.  In this case also the missing transverse mass variable ($M^{T}_{\gamma\subDP}$) [see Fig.~\ref{fig:transv_mass} (right plot)] turns out to be particularly useful to suppress the SM backgrounds.
Assuming an integrated luminosity of 100 fb$^{-1}$ the 
5$\sigma$ reach in branching ratio is  about BR$_{\gamma \subDP}\!\!\simeq$ 2\%. 
With the HL-LHC integrated luminosity of 3 ab$^{-1}$, the 5$\sigma$ reach can be extended down to BR$_{\gamma \subDP}\!=3.4\times 10^{-3}$.
A comparison of significances for both of these Higgs production channel is presented
in Table~\ref{tab:signf}~\cite{Biswas:2016jsh,Biswas:2017anm}.

\begin{table}
\begin{center}
\begin{tabular}{c|c c|c c|c c}
 BR$_{\gamma \subDP}$ (\%)  &\multicolumn{2}{|c|}{\;L=100\,fb$^{-1}$} & \multicolumn{2}{|c|}{\;L=300\,fb$^{-1}$} &\multicolumn{2}{|c}{\;L=3\,ab$^{-1}$} \\ \hline
{\small Significance} & 2$\sigma$ & 5$\sigma$ & 2$\sigma$ & 5$\sigma$ & 2$\sigma$ & 5$\sigma$ \\ \hline
\;BR$_{\gamma \subDP}$(VBF) &  0.76 & 1.9 &  0.43 & 1.1  &  0.14 & 0.34 \\ \hline
BR$_{\gamma \subDP}\,$($gg$F) & \, 0.064 &  0.16   & \, 0.037  &  0.092  &\,  0.012  &  0.029  \\ \hline
\end{tabular}
\end{center}
\caption{Reach in BR$_{\gamma \subDP}$ (in percentage)  
for a 95\% C.L.
(2$\sigma$)  exclusion or a 5$ \sigma$ discovery at the 14 TeV LHC, in the VBF and 
gluon-fusion channels, for different integrated luminosities~L~\cite{Biswas:2016jsh,Biswas:2017anm}.}
\label{tab:signf}
\end{table}

\begin{table}[t] 
\begin{center}
\begin{tabular}{l|c|c|c|c}
Cuts                           & Signal   &  $\gamma+$jets   &  $\gamma+Z+$jets  &  QCD multiijet  \\ \hline
Basic cuts                    &  17.7    &  266636               &              1211            &   72219            \\
Rapidity cuts                &    8.8     &   8130                  &               38.1            &   33022             \\
$M^{T}_{\gamma\subDP}$ cuts         &    5.0     &     574                  &                6.5             &     3236             \\
\hline
\end{tabular}
\end{center}
\caption{Cross sections  times acceptance $\sigma\times A$ (in fb) for  the VBF signal and backgrounds at 
14~TeV, after sequential application of cuts defined in the text, assuming BR$_{\gamma \subDP}$=1\%~\cite{Biswas:2016jsh}.}
\label{vbf1}
\end{table}


\section{Experimental searches at the LHC}

In this section, we summarize present experimental LHC results on dark-photon production via Higgs decay in two different Higgs production channels, namely, VBF and Higgs production in association with a $Z$-boson ($ZH$).

\subsection{VBF production}
Both the ATLAS and CMS experiments studied the $pp \rightarrow \gamma + \slashed{E}_T + jets$  
via Higgs production
in VBF channel \cite{ATLAS:2021pdg,CMS:2020krr}. The data collected in this channel has been
interpreted in the
context of Higgs production through VBF and its subsequent decay to a photon and 
massless dark photon which goes undetected, $ pp \to Hjj\rightarrow\gamma\DP jj$.

The SM backgrounds for this process are $V\gamma +jets$, where $V=Z,W$ and $\gamma +jets$.  In case of $Z\gamma +jets$ the $Z$ decays to a pair of $\nu\bar{\nu}$ and gives rise to isolated photon, missing transverse momentum and jets. $W\gamma +jets$
contributes to the same final state when the $W^{\pm}$ boson decays to a lepton and a neutrino and the lepton goes missing as it may not satisfy the required identification 
criteria. One can also have contribution from $W\gamma +jets$ and $W\gamma +jets$ process with a jet being mistagged as photon.

\subsubsection{ATLAS}

The ATLAS experiment at the LHC is a particle detector having a cylindrical geometry
with forward-backward symmetry. It consists of mainly four parts: the inner most
tracking detector, electromagnetic and hadronic calorimeter (ECAL and HCAL), and the outer muon spectrometer. The tracking detector which is used to measure the momentum of charged particles has a rapidity coverage of $|\eta|<2.5$. The ECAL and HCAL coverage is up to $|\eta|<4.9$. The inner tracking detector is provided with a 2.0 T axial magnetic field produced by a surrounding superconducting solenoid. The muon spectrometer is based on large superconducting toroidal magnets and provides an integral field in the range 2.0 T to 6.0 T. 

The ATLAS analysis of $H\to \gamma\DP$ in the VBF channel corresponds to $139$ fb$^{-1}$ data collected by the 
ATLAS collaboration during 2015-2018 at 13 TeV LHC collision energy~\cite{ATLAS:2021pdg}. ATLAS experiment uses several kinematic variables similar to those
discussed in Section 3.2 in addition to their dedicated object reconstruction criteria.

The results of ATLAS analysis can be used to set limits on the cross section
of Higgs production in the VBF channel times BR($H\to \gamma\DP$) as a function of the hypothetical neutral Higgs boson in the mass range $60 ~{\rm GeV} < m_H < 2$ TeV (Fig.~\ref{fig:ATLAS2}). The corresponding 
bound obtained by the ATLAS experiment is $0.19$ pb. Assuming
a SM-like Higgs production cross section in the VBF channel in this mass range the results can be interpreted as a bound on BR($H\to \gamma\DP$). For the SM Higgs boson ($m_H\simeq125$ GeV) the 
95\% C.L. upper bound on BR($H\to \gamma\DP$) obtained by the ATLAS collaboration corresponds to $0.014$.

\begin{figure}[h] 
\centering
\includegraphics[width=0.6\textwidth]{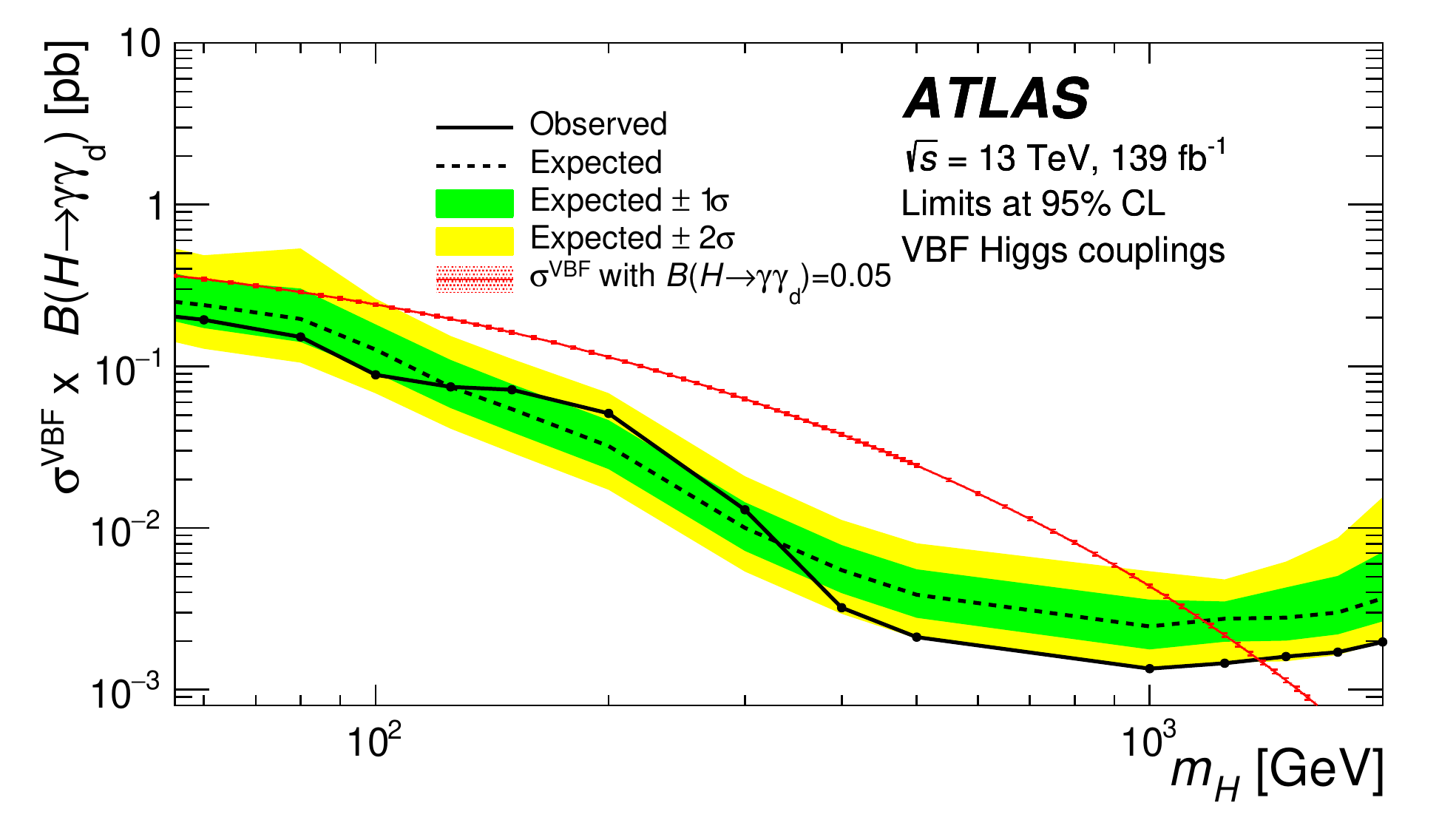}
\caption{
The observed and expected $95\%$ C.L. upper limit on  the Higgs production cross section times $\BR(H \to  \gamma \DP)$, for various scalar mass hypothesis~\cite{ATLAS:2021pdg}. The red line  corresponds to the theoretical SM-like Higgs production cross section in VBF channel times BR($H\to \gamma\DP$)$\sim 5\%$.}
\label{fig:ATLAS2}
\end{figure} 

\subsubsection{CMS}
 The CMS detector also consists of an inner tracker, electromagnetic calorimeter, hadronic calorimeter, and muon detector. The inner tracker operates in the range $|\eta| <2.5$. The ECAL and HCAL has rapidity coverage of $|\eta| <3.0$. In addition, the forward calorimeter provides a rapidity coverage up to $|\eta| <5.0$. CMS also has a dedicated muon detector which constitute the outer most layer of the CMS detector.
 
The results of the corresponding CMS analysis in the VBF channel is shown in Fig.~\ref{fig:cmslimits1} which corresponds to 130 fb$^{-1}$ of data
collected by CMS collaboration during 2016-2018 at 13~TeV LHC collision energy
\cite{CMS:2020krr}. The observed 95\% C.L. on the $H\to \gamma\DP$ branching ratio obtained by CMS collaboration in the VBF channel is 3.5\% for the SM Higgs boson with $m_H=125$~GeV \cite{CMS:2020krr}.

\begin{figure}[hbt]
\centering
\includegraphics[width=0.5\textwidth]{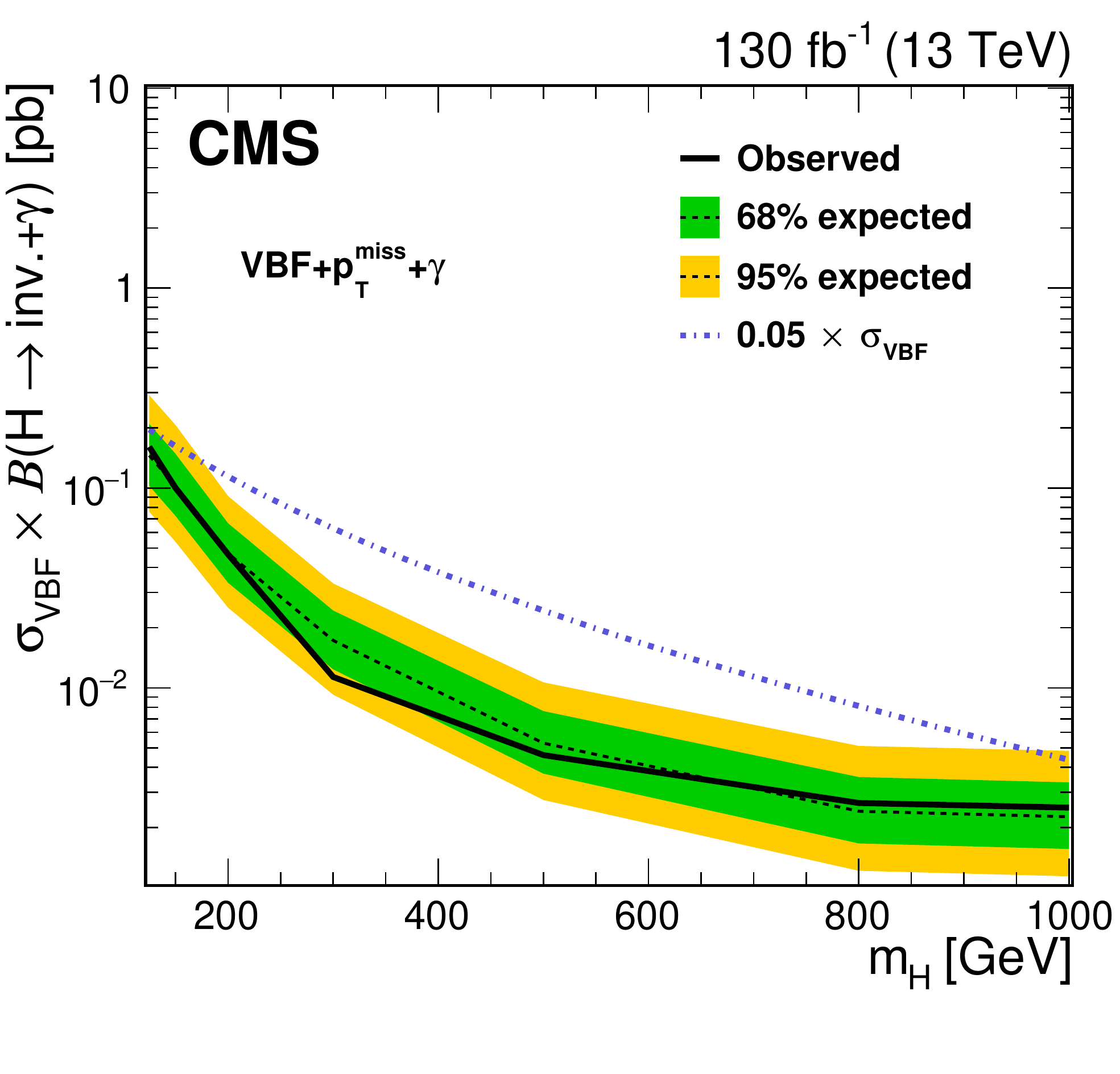}
\caption{
  CMS bound on the cross section ($\sigma_{VBF}$) times $\BR(H \to \text{inv.}+\gamma)$
 in the VBF channel  as a function of the scalar mass ($m_H$)~\cite{CMS:2020krr}. 
 The solid black line corresponds to the observed limit while the black dashed line corresponds to the expected limit, at 95\% C.L. The dot-dashed line represents the signal 
corresponding to a SM $\sigma_{\rm VBF}$ value and $\BR(H \to \text{inv.}+\gamma)=5\%$.}
   \label{fig:cmslimits1}
\end{figure}

\subsection{ZH production}

\subsubsection{CMS}

The CMS collaboration has also studied the Higgs production in association with a $Z$-boson in $pp$ collisions with subsequent decay of the Higgs into a photon plus an undetected particle using the $137$ fb$^{-1}$ of data 
collected at 13-TeV $pp$ c.m. energy \cite{CMS:2019ajt}. In the absence of any significant excess over the SM backgrounds an exclusion limit can be set on theoretical models predicting such exotic decay of the Higgs boson. The results of CMS
study has been interpreted in the context of models predicting $H\to \gamma\DP$ decay and the corresponding process $pp \to ZH \to  (Z\to \ell^-\ell^+)(H\to \gamma\DP)$. The Feynman diagram for the above process is shown in Fig.~\ref{fig:FD-LHC-ZH}.

\begin{figure}[t!]
  \begin{center} 
  \includegraphics[width=3in]{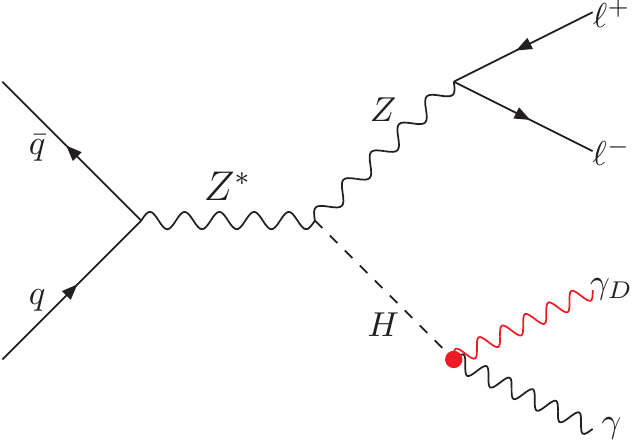}
  \caption{{\small Feynman diagrams for the Higgs production in association with $Z$-boson and the subsequent decay $H\to \gamma\DP$ at the LHC.
}}
\label{fig:FD-LHC-ZH}
\end{center}
\end{figure}

The main backgrounds to the final state under consideration as analyzed by CMS are $ZW$, $ZZ$ and $Z\gamma$. For  $ZW$ and $ZZ$, the contribution comes
when a lepton is misidentified as a photon. 
 
The results of the CMS analysis in the $ZH$ channel is presented in Fig.~\ref{fig:cmslimits2}, which provides an exclusion limit on the cross section times branching ratio of the Higgs
in the {\it photon+invisible} mode~\cite{CMS:2019ajt}. If the data is interpreted in the context of theoretical models predicting a $H\to \gamma\DP$ decay, a bound on cross section times BR($H\to \gamma\DP$) $\sim 0.04$ pb can
be obtained for $m_H\simeq125$~GeV. The observed 95\% C.L. upper bound on BR($H\to \gamma\DP$)  obtained by CMS  is 4.6\% for the SM Higgs boson with $m_H\simeq125$ GeV.

\begin{figure}[hbtp]
\centering
\includegraphics[width=0.5\textwidth]{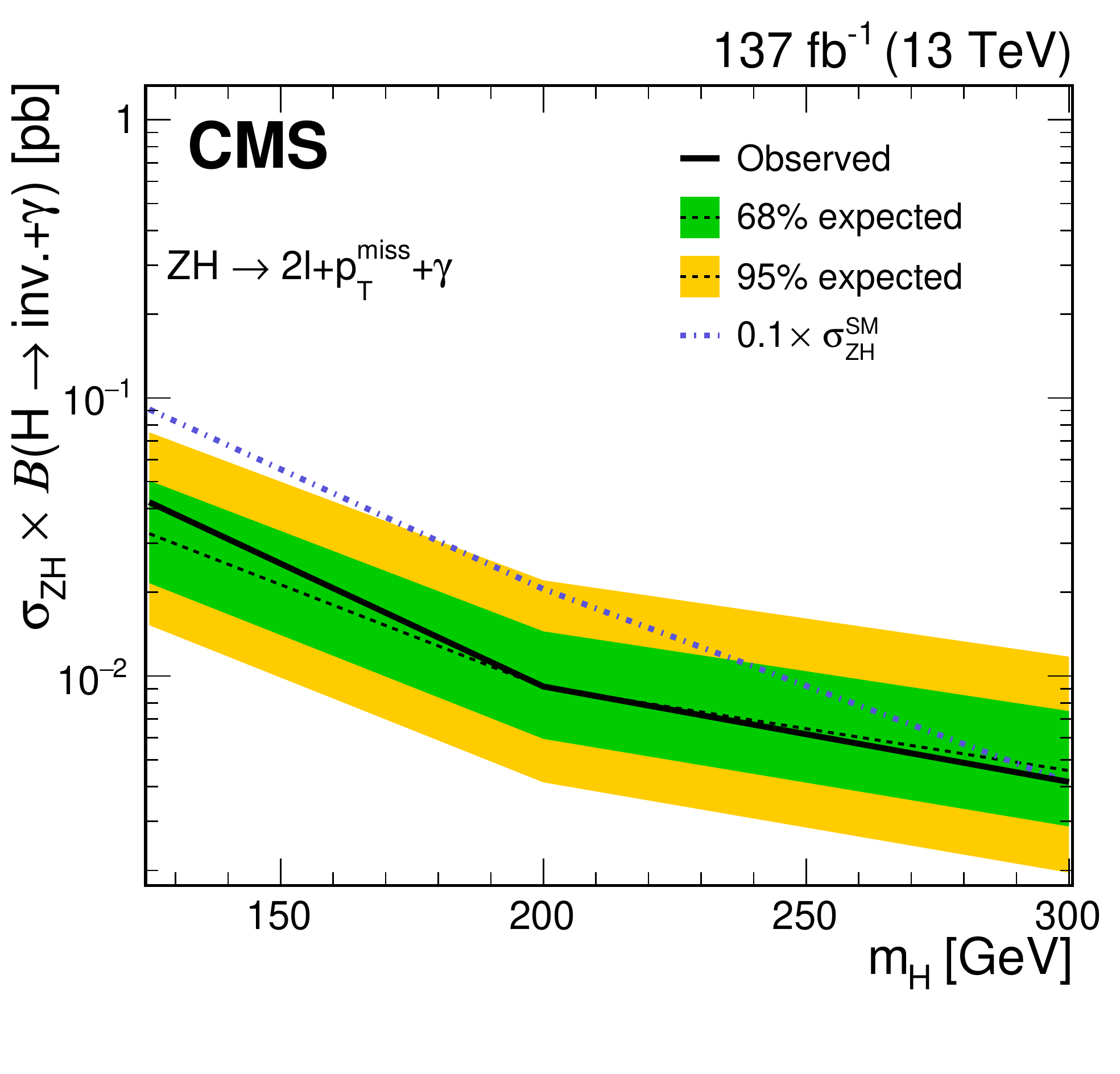}
\caption{
CMS bound on the cross section ($\sigma_{ZH}$) times $\BR(H \to \text{inv.}+\gamma)$ in the ZH channel as a function of the scalar mass ($m_H$)~\cite{CMS:2019ajt}. 
The solid black line corresponds to the observed limit while the black dashed line corresponds to the expected limit, at 95\% C.L.. The dot-dashed line represents the signal 
corresponding to a SM  $\sigma_{ZH}$ value and $\BR(H \to \text{inv.}+\gamma)=10\%$.}
   \label{fig:cmslimits2}
\end{figure}

The CMS collaboration also extended the analysis for a heavier hypothetical neutral scalar boson in the mass range 125 GeV to 300 GeV with similar decay mode.

The CMS bound from $ZH$, $VBF$ and combined analysis are summarized in Table~\ref{tab:comblimits}, for the SM Higgs boson with $m_H\simeq125$ GeV.

\begin{table}[htb]
  \centering
  \begin{tabular} {cccccc}
\hline
\multicolumn{2}{c}{VBF} & \multicolumn{2}{c}{$ZH$} & \multicolumn{2}{c}{VBF+$ZH$} \\
Obs. (\%) & Exp. (\%) & Obs. (\%) & Exp. (\%) & Obs. (\%) & Exp. (\%) \\
\hline
3.5 & $2.8^{+1.3}_{-0.8}$ &  4.6 & $ 3.6^{+2.0}_{-1.2}$ & 2.9  & $2.1^{+1.0}_{-0.7}$ \\
  \hline
  \end{tabular}
    \caption{Observed and expected 95\% C.L. upper limits  on BR($H\to\text{inv.}+\gamma$) at $m_{H}\simeq 125$ GeV, 
  from the VBF, $ZH$ channels, and  the combined analysis~\cite{CMS:2020krr}.}
  \label{tab:comblimits}
\end{table}


\subsection{Future perspectives at the LHC}

The future prospects of dark-photon searches at the HL-LHC  via Higgs production in $pp$ collisions 
are summarized in Table~\ref{tab:HE-HL-LHC_reach}  [including also the possibility of a High Energy (HE) LHC at $\sqrt{s}=27$ TeV]. Using a similar analysis as that implemented by CMS at $\sqrt{S}=8$ TeV 
 which can provide a good control over the overwhelming $\gamma+jets$ and $jj$ backgrounds (called ``CMS inspired" in Table~\ref{tab:HE-HL-LHC_reach}), one can achieve strongest limit on
$\BR(H\rightarrow \gamma\DP)$. At a c.m. energy $14$ ($27$) TeV with integrated luminosity of $3$ ($15$) ab$^{-1}$, the expected $5\sigma$ 
discovery reach on the BR($H\rightarrow \gamma\DP$) is $\sim 3\times 10^{-4}$ ($1.3 \times 10^{-4}$), while the corresponding expected $2\sigma$ exclusion limit on
the Higgs branching ratio in the same mode is found to be \mbox{$\sim 1.2\times 10^{-4}$ ( $0.5 \times 10^{-4}$)}~\cite{CidVidal:2018eel}.

\begin{table}[h!]
\begin{center}
\small\begin{tabular}{c|c c|c c}
$\BR_{\gamma \subDP}(\%)$&\multicolumn{2}{|c|}{$3 ~{\rm ab}^{-1}$@14 TeV}&\multicolumn{2}{|c}{$15 ~{\rm ab}^{-1}$@27 TeV} \\ \hline
significance & $2\sigma$ & $5\sigma$ & $2\sigma$ & $5\sigma$ \\ \hline
CMS inspired & $0.012$  & $0.030$ & $0.0052$ & $0.013$ \\ \hline
\end{tabular}\normalsize
\caption{The future projection at the HL-LHC and HE-LHC in terms of discovery ($5\sigma$) reach and exclusion ($2\sigma$) limit for  BR($H\rightarrow \gamma\DP$) (in \%)~\cite{Biswas:2017anm,CidVidal:2018eel}.}
\label{tab:HE-HL-LHC_reach}
\end{center}
\end{table}


\section{Dark photon production at  future $e^+e^-$ colliders}

It is also important to look at future prospects for dark-photon searches via Higgs-production in the context of various proposed future $e^+e^-$ colliders. 
In~\cite{Biswas:2015sha}, we have shown 
that one could improve the sensitivity to the $H\to \gamma\DP$ branching ratio at future $e^+e^-$ collider experiments. In particular, the proposed Future Circular $e^+e^-$ Collider (${\rm FCC{ ee}}$) is deemed to run  with high luminosity at c.m. energies $[91.2, 161, 240, 350(365)]$~GeV, corresponding, respectively, to the $Z$ pole and to the approximate $WW$, $ZH$ 
and $t\bar{t}$ thresholds~\cite{FCC:2018evy}. 
We have proposed both direct dark-photon production in association 
with a Higgs boson ($e^+e^-\to H\DP$)~\cite{Biswas:2015sha}, and the dark-photon production in the decay of a Higgs boson [$e^+e^-\to ZH \to  Z(H\to \gamma\DP)$]~\cite{Biswas:2017lyg}, by focusing  
on the c.m. energy  $\sqrt{s} \simeq 240\ {\rm GeV}$ with integrated luminosity of $10\ {\rm ab}^{-1}$.

In the context of $e^+e^−$ colliders all the signal and
backgrounds events are generated using Madgraph event generator and
then interfaced with PYTHIA for further analysis. However, to simulate
the process $e^+e^− \to H\DP$ using Madgraph we have to implement the
appropriate effective operators $(F^{\mu\nu}F^D_{\mu\nu}H)$, $(Z^{\mu\nu}F^D_{\mu\nu}H)$  in the Madgraph model file. This has been accomplished with the help
of FeynRules (v2.0) \cite{Alloul:2013bka}, where we have implemented these operators at the Lagrangian level and the corresponding output of the FeynRules are then interfaced with Madgraph.

\begin{itemize}

\item \noindent The study of $e^+e^-\to H\DP \to (H\to b\bar{b})\DP$ illustrates a novel signature in which an invisible massless system recoils against a $b\bar{b}$ system with invariant 
mass close to the Higgs mass. This is a very unique feature of a massless dark-photon produced in association with a Higgs. The corresponding Feynman diagram is shown in Fig.~\ref{fig:FD-ee_HDPH}. 
Due to the clean environment in a $e^+e^-$ collider and definite 
knowledge of the initial state, one can in principle reconstruct the full four-momentum of the invisible dark-photon system. The SM backgrounds to the $b\bar{b}+{\rm \it missing ~energy}$
final states are $\nu\bar{\nu} b\bar{b}$ and $\nu\bar{\nu} q\bar{q}$. The contribution to the missing energy ($\slashed{E}$) in the background process is due to the pair of invisible almost massless neutrinos.
Here, missing energy is defined as ${\rm \slashed{E}}=\sqrt s-\sum {\rm E}_{visible}$ where the sum is over all the visible particles.
In \cite{Biswas:2015sha} we have pointed out  how introducing various kinematic variables such as invariant mass of the two leading jets ($M_{jj}$), missing energy ($\slashed{E}$), and missing mass
($M_{\rm miss}$) one can efficiently suppress the SM background.  The variable missing mass is defined as $M_{\rm miss}=\sqrt{\slashed{E}^2-\slashed{\bf p}^2}$ where ${\rm \slashed{\bf p}}=-\sum {\mathbf p}_{visible}$ 
is the final-state missing three-momentum vector.  The missing mass plays a crucial role to separate the signal from SM backgrounds. For the signal process the corresponding distribution is centered around $M_{\rm miss} = 0$
due to the presence of single massless dark-photon in the final state [see Fig.~\ref{fig:missing_mass} (left)]. 

\begin{figure}[t!]
  \begin{center}
  \includegraphics[width=3in]{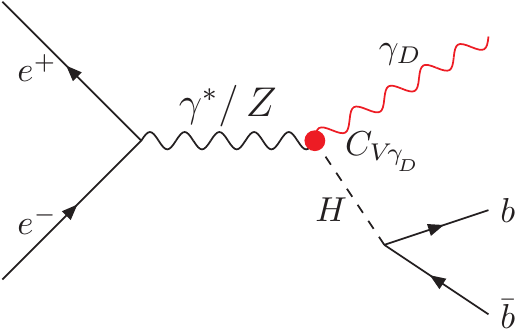}
  \caption{{\small Feynman diagrams for the $H\DP$ production  at  $e^+e^-$ colliders.
}}
\label{fig:FD-ee_HDPH}
\end{center}
\end{figure}

\begin{figure}
\begin{center}
\includegraphics[width=0.4\textwidth]{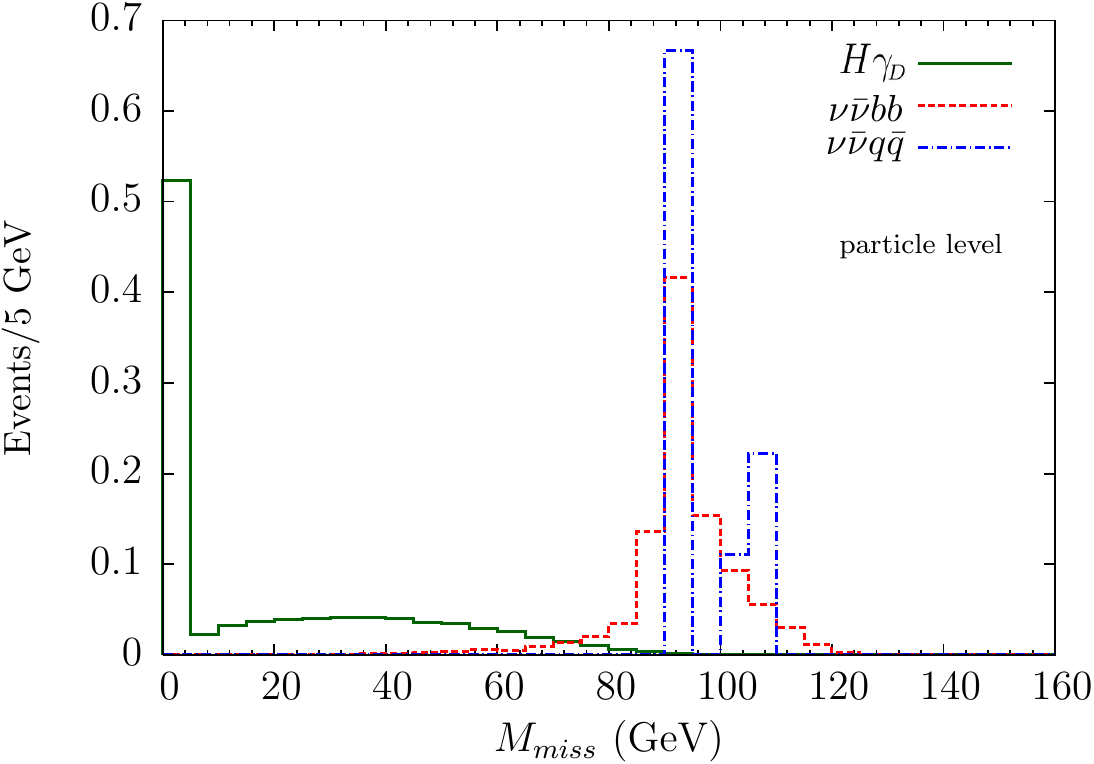}
\hskip 20pt \includegraphics[width=0.4\textwidth]{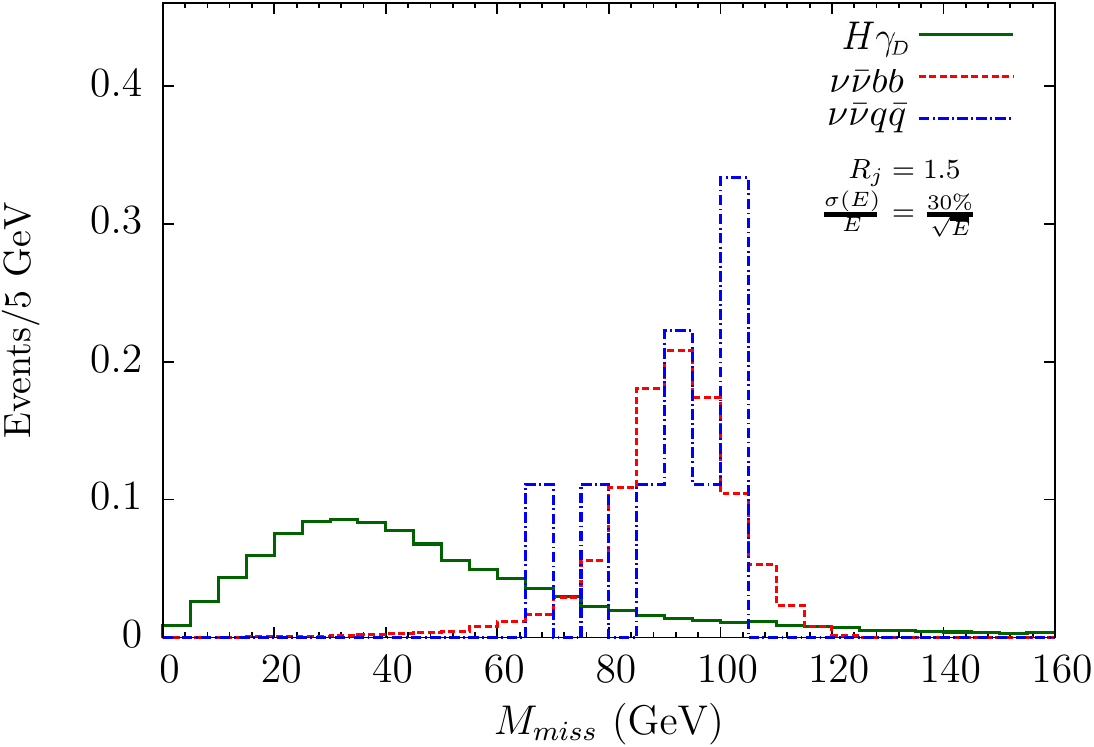}
\caption{Missing mass  distribution without ({\it left}) and with ({\it right}) taking into account the effects of finite detector resolution in the $e^+e^-\to H\DP \to (H\to b\bar{b})\DP$ channel~\cite{Biswas:2015sha}.}
\label{fig:missing_mass}
\end{center}
\end{figure}

In Table~\ref{table:acceptancies}, we summarize the cross sections and cut efficiencies for the signal and background processes. For detailed event selection criteria see~\cite{Biswas:2015sha}. 
The corresponding sensitivity reach as a function of $C_{\gamma\subDP}$  and $C_{Z\subDP}$ are shown in Fig.~\ref{fig:significance}, which illustrates that, at 95\% C.L., 
one can exclude $C_{\gamma\subDP}>1.9$ (for $C_{Z\subDP}=0$). This can be translated to an exclusion limit on the BR($H\to \gamma \DP $) greater than 3 times the SM BR($H\to \gamma \gamma$).
For $C_{\gamma\subDP}=0$ the corresponding exclusion limit is $C_{Z\subDP}>2.7$, and for $C_{Z\subDP}\simeq0.79\;C_{\gamma\subDP}$ the exclusion limit is $C_{\gamma\subDP}>1.6$.

\begin{table}
\begin{center}
\begin{tabular}{||l||c||c||}
\hline 
\hline
Process & Cross section (fb) & Acceptance after cuts (\%) \\
\hline 
$ H \DP$\;\; ($C_{Z \subDP}= 0$)   &  $10.1\times 10^{-3} \;C^2_{\gamma \subDP}$   & 17.3  \\
\hline
$ H \DP$\;\; ($C_{\gamma \subDP}= 0$)     &  $4.8\times 10^{-3}\;C^2_{Z \subDP}$   & 17.3  \\
\hline
$ H \DP$\;\; ($C_{Z \subDP}=0.79 \;C_{\gamma \subDP}$)   &  $13.8\times 10^{-3} \;C^2_{\gamma \subDP}$   & 17.3 \\
\hline
SM $\;\nu\bar{\nu}b\bar{b}$   &  115.   &  0.08  \\
\hline
\hline
\end{tabular}
\caption{ Cross sections (in fb) and corresponding acceptances after kinematical  cuts on signal and SM backgrounds at  \mbox{$\sqrt s=$240 GeV~\cite{Biswas:2015sha}.}  Applied cuts include basic cuts for object reconstruction, 
dijet invariant mass ($M_{jj}$) to be within 10\% of the $M_{jj}$ peak value  of  signal events, \mbox{$M_{\rm miss}<40$ GeV}, and \mbox{$\slashed E < 100$ GeV. Cross sections include BR$(H\to
b\bar{b})\simeq0.58$.}} 
\label{table:acceptancies}
\end{center}
\end{table}     

\begin{figure}
\begin{center}
\includegraphics[width=0.5\textwidth]{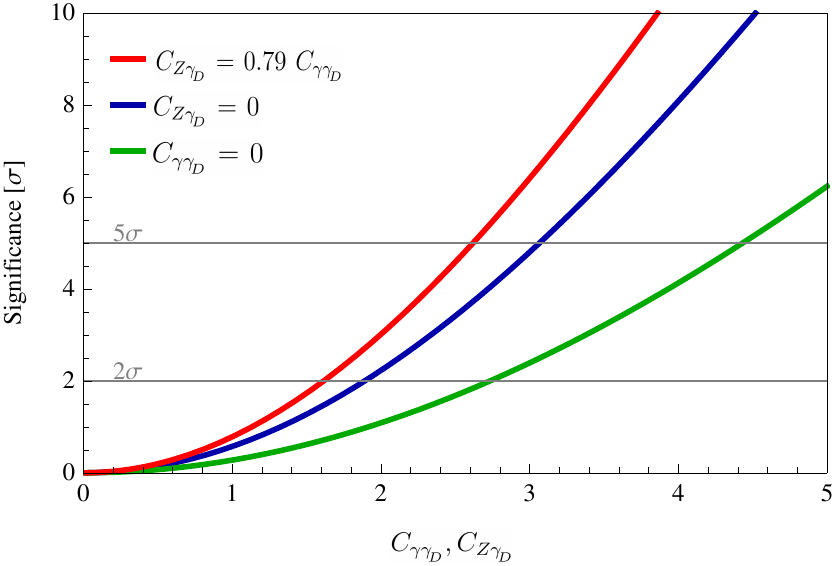}
\caption{Estimated signal significance ($S/\sqrt{S+B}$) as a function of the effective couplings
$C_{\gamma\subDP}, C_{Z\subDP}$ for $e^+e^- \to H\DP$ channel 
at a center of  mass energy 240 GeV and integrated luminosity of 10 ab$^{-1}$~\cite{Biswas:2015sha}. The solid green line represents the signal significance for $C_{\gamma\subDP}= 0$, 
solid blue line represents the same for $C_{Z\subDP}= 0$, while the solid red line corresponds to the case $C_{Z\subDP}= 0.79C_{\gamma\subDP}$.  The 5$\sigma$ discovery reach 
and 2$\sigma$ exclusion limit are shown by the upper and lower horizontal gray lines, respectively.}
\label{fig:significance}
\end{center}
\end{figure}

\item \noindent  The dark-photon production via Higgs decay at the future $e^+e^-$ collider, $e^+e^-\to ZH \to  Z(H\to \gamma\DP)$ provides a better sensitivity to the $H\to \gamma\DP$ 
branching ratio. We have proposed two different final states considering both leptonic and hadronic decay modes of the $Z$-boson. The leptonic final state in $e^+e^-\to ZH \to  (Z\to \mu^-\mu^+)(H\to \gamma\DP)$
consists of a pair of opposite sign muons, an isolated photon and missing energy due to the presence of a massless invisible dark-photon. In the hadronic final states the muon pair is replaced by a pair of jets in $e^+e^-\to ZH \to  (Z\to q\bar{q})(H\to \gamma\DP)$. 
The corresponding SM background contributions dominantly come from $e^+e^-\to ZH \to  Z(H\to \gamma{\gamma})$ and $e^+e^- \to  Z \gamma{\gamma}$ when one of the photons is not detected at the detector and/or lies in the forward region. 
The Feynman diagram(s) for this process is illustrated in Fig.~\ref{fig:FD-ee_ZH}
The kinematic variables such as missing energy ($\slashed{E}$), and missing mass
($M_{\rm miss}$) proposed in the previous analysis along with dimuon/dijet invariant mass variable ($M_{\mu^+\mu^-/jj}$) help to reduce the contribution of these SM backgrounds. In addition, one has the advantage of using invariant mass of the {\it photon+dark-photon system} ($M_{ \gamma\subDP}$)
to further discriminate the signal from backgrounds, thanks 
to the full reconstruction of the dark-photon momenta in a $e^+e^-$-colliding environment. For the signal process the last three kinematic distributions are centered around $ M_{\rm miss}= 0$,  $M_{\mu^+\mu^-/jj} = M_Z$,  and $M_{\gamma\subDP} = m_H$, respectively. In Tables~\ref{muon channel cut flow} 
and ~\ref{hadron channel cut flow} , we summarize 
the cut-flow effects   for both the dimuon and dijet ($+\gamma+\slashed{E}$) final states after imposing a set of event selection criteria detailed in~\cite{Biswas:2017lyg}.

\begin{figure}[t!]
  \begin{center}
  \includegraphics[width=2.0in]{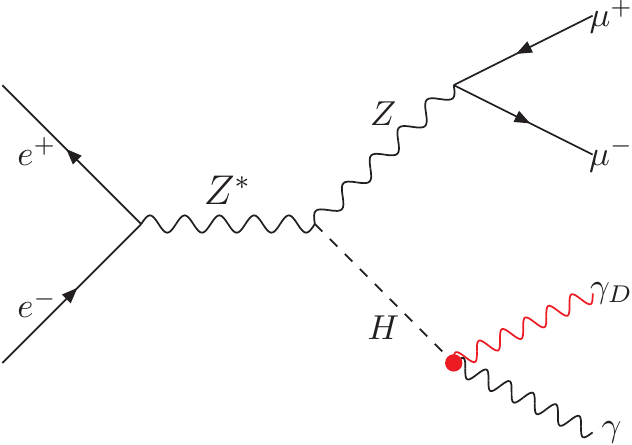}
  \hskip 30pt \includegraphics[width=2.0in]{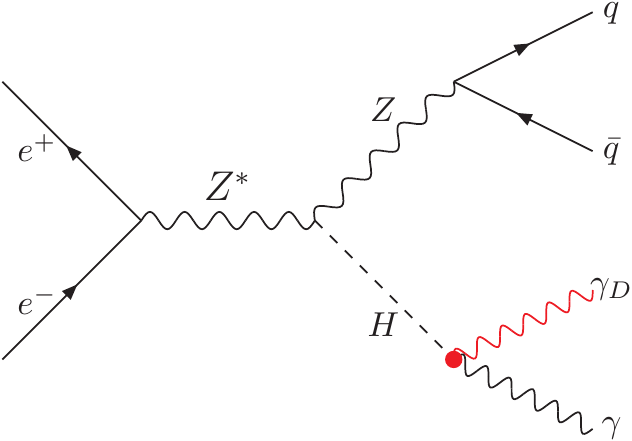}
  \caption{{\small Feynman diagrams for dark photon production in $e^+e^-$ collisions via
       associated $ZH$ production.
}}
\label{fig:FD-ee_ZH}
\end{center}
\end{figure}

\begin{table}
\begin{center}
\begin{tabular}{|l||c|c|c|c|}
\hline
Process & Basic cuts  & $M_{\ell\ell}$ cut & $M_{\gamma\subDP}$ cut & $M_{\rm miss}$ cut \\
\hline
$\mu^+\mu^- \gamma\DP$ \;\;($\BR_{\gamma\subDP} = 0.1\%$) & 65.3 & 54.9 & 49.7 & 47.3 \\
\hline
$\mu^+\mu^-\nu\bar{\nu}\gamma\;\;$  & $5.00\times 10^{4} $& $5.73 \times 10^{3}$& $1.09 \times 10^{3}$& 15 \\
\hline
\end{tabular}
\caption{Event yields after sequential cuts for $e^+e^-\!\!\rightarrow ZH\rightarrow \mu^+\mu^- \gamma\DP$ and corresponding background, for an integrated luminosity of $10\ {\rm ab}^{-1}$, and  c.m. energy  $\sqrt{s} = 240\ {\rm GeV}$~\cite{Biswas:2017lyg}. The signal yield has been normalised assuming BR($H \to {\gamma\DP}$)$=0.1\%$.} 
\label{muon channel cut flow}
\end{center}
\end{table} 

\begin{table}
\begin{center}
\begin{tabular}{|l||c|c|c|c|c|}
\hline
Process & Basic cuts &  $M_{jj}$ cut & $M_{\gamma\subDP}$ cut & $M_{\rm miss}$ cut & $\slashed{E}$ cut \\
\hline
$jj  \gamma\DP$\;\; ($\BR_{\gamma\subDP} = 0.1\%$) & 804 & 669 & 154 & 110 & 72 \\
\hline
$jj\gamma$ & $3.39\times 10^{7}$ & $2.26\times 10^{7}$ & $1.47\times 10^{5}$ & $6.5\times 10^{4}$ & -- \\
\hline
 $jj\nu\bar{\nu}\gamma$ &  $3.9\times 10^{4}$  &  $3.1\times 10^{4}$   &  $5.9\times 10^{3}$   &  2.2   &  --  \\
\hline
\end{tabular}
\caption{Event yields  for $e^+e^-\!\!\rightarrow ZH\rightarrow q\bar q \gamma\DP$, 
  after sequential cuts discussed in~\cite{Biswas:2017lyg}, and corresponding backgrounds rates, for an integrated luminosity of $10\ {\rm ab}^{-1}$, and  c.m. energy  $\sqrt{s} = 240\ {\rm GeV}$. The signal yield has been normalised
  assuming BR($H \to \gamma\DP$)$=0.1\%$. Dashes stand for event yields less than 1.} 
\label{hadron channel cut flow}
\end{center}
\end{table}     

The sensitivity reach as a function of BR($H \to {\gamma\DP}$) is depicted in Fig.~\ref{combined significance} for an integrated luminosity of $10$ ab$^{-1}$ at $\sqrt s=240$ GeV. Clearly, the hadronic channel provides a better discovery reach (or exclusion limit) compared to the 
dimuon channel.  The combined $5\sigma$ sensitivity for discovery reaches  BR($H \to {\gamma\DP}$) $\simeq 2.7\times 10^{-4}$, while the 95\%~C.L. exclusion limit  is again BR($H \to {\gamma\DP}$) $\simeq 0.5\times 10^{-4}$.
 
\begin{figure}[h!]
\begin{center}
\includegraphics[width = 0.7\textwidth]{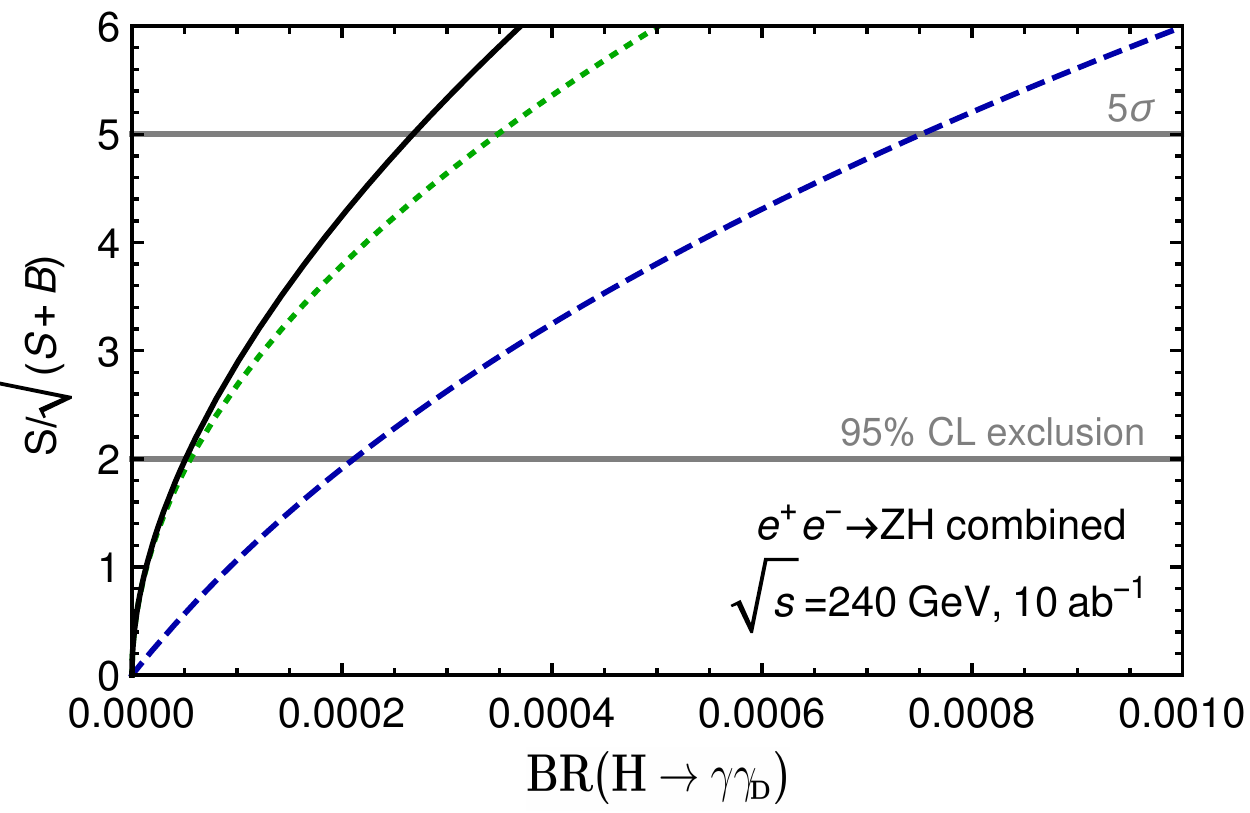}
\caption{Estimated signal significances vs the branching ratio of $H \to \gamma\DP$ for $e^+e^- \to ZH$ channel at a center of  mass energy 
240 GeV and integrated luminosity of 10 ab$^{-1}$~\cite{Biswas:2017lyg}. The blue dashed line corresponds to the estimated significance in the dimuon+$\gamma+\slashed{E}$ 
final state, the green dotted line corresponds to that in the dijet+$\gamma+\slashed{E}$ final state, while the significance in the combined channel is represented 
by the solid black line. The 5$\sigma$ discovery reach and 2$\sigma$ exclusion limit are shown by the upper and lower horizontal gray lines, respectively.}
\label{combined significance}
\end{center}
\end{figure}

\end{itemize}

\section{Conclusions}
We have explored Higgs-mediated dark photon production at the LHC and future colliders via the  Higgs decay into a photon and a dark photon $H\to \gamma \DP$. We have assumed the dark photon to be massless and associated to an unbroken $U(1)_D$ gauge symmetry in the dark sector. Contrary to the massive case, a massless dark photon is only coupled to the SM fields via higher dimensional operators, which are suppressed by the unknown UV scale $\Lambda$. Then, depending on $\Lambda$, dark photon direct production at colliders is in general strongly suppressed by terms of order ${\cal O}(E/\Lambda)^2$, with $E$ the characteristic energy of the process, thus rendering the search for a massless dark photon dramatically dependent on the UV completion of the theory. Nevertheless, due to its non-decoupling properties, the Higgs boson system 
contradicts this expectation, and offers a privileged strategy to explore the dark photon production, by a clean experimental probe. 

By using a simplified model with UV completion,  we have shown that the  $H\to \gamma \DP$ decay might have measurable rates (mostly insensitive to the UV scale) which depend only on a few dimensionless (and potentially large) dark-sector parameters.
We have assumed the existence of a set of scalar messenger fields  connecting the SM  and dark-sector fields via renormalizable interactions. We have provided  correlated predictions for the  $H\to \gamma \DP,  \DP \DP$ decay rates, and NP exotic contributions to the SM channels $H\to \gamma \gamma, Z \gamma, g g$, from which  the non-decoupling properties clearly emerge. Analytical results for the corresponding amplitudes and decay rates  are provided in the most general scenario, with generic $N$ scalar messenger fields  charged under both SM  and  $U(1)_D$
gauge interactions.

Since a massless or ultralight $\DP$ would be experimentally invisible, the  typical signature  for the $H\to \gamma \DP$ decay at colliders would be characterized by an almost monochromatic photon -- with energy half of the Higgs boson mass in its rest frame -- plus missing transverse energy. We review the main signatures related to the $H\to \gamma \DP$ decay in the dominant Higgs production channels at the LHC, and discuss the most relevant irreducible and reducible backgrounds. In particular, the $H\to \gamma \DP$ decay in the gluon-gluon fusion and VBF mechanisms (analyzed in \cite{Gabrielli:2014oya,Biswas:2016jsh})
are discussed and compared with the recent analysis by the ATLAS \cite{ATLAS:2021pdg} and CMS \cite{CMS:2020krr,CMS:2019ajt} 
collaborations  for the search of a $H\to \gamma \DP$ signal at the LHC, in  VBF and associated-$ZH$ events collected at  $\sqrt{s}\simeq13$ TeV, with  about 140${\rm fb}^{-1}$. No significant excess above the SM expectations is found in either experiments, leading to an observed  95\% C.L. upper limit on  $\BR(H\to \gamma \DP)$ of 1.4\% (ATLAS)  \cite{ATLAS:2021pdg} and 2.9\% (CMS) \cite{CMS:2020krr}.

These results are compared with the corresponding predictions for the $\BR(H\to \gamma \DP)$ in the UV complete model for the dark sector, where a  model independent parametrization for the $H\to \gamma \DP,  \DP \DP$ decay rates  has been adopted. In particular,  $\BR(H\to \gamma \DP)$ regions allowed by all present constraints, as a function of the $\alpha_D$ -- the fine-structure constant related to the $U(1)_D$ gauge symmetry -- have been presented. We have found that the current sensitivity in the $\BR(H\to \gamma \DP)$ measurements by ATLAS and CMS, which is at the percent level, is presently one order of magnitude weaker than what is needed for detecting $\BR(H\to \gamma \DP)$ in the allowed range, consistent with actual constraints on $\BR(H\to \gamma \gamma)$ and  $\BR(H\to {\rm invisible})$. Hence a larger statistics will be needed
at the LHC in order to explore the allowed $\BR(H\to \gamma \DP)$ range at the permil level.
Future perspectives for the search of the $H\to \gamma \DP$ signal at future $e^+e^-$ colliders and hadron colliders experiments are also shown.

We also discussed possible alternative new physics scenarios that could fake the dark-photon signature, by analyzing the generic $H\to \gamma X$ decay, with $X$ an invisible (light) dark particle. We have shown that the observation of the monochromatic photon signature plus missing energy   identifies the dark photon as {\it by far} the most viable interpretation. 
Indeed,  we have shown that, although   both scalar and pseudoscalar $X$ cases are forbidden by angular momentum conservation,  bosonic $X$ particles with spin higher than 1 are in principle possible (while fermionic $X$ states are forbidden by Lorentz invariance). In particular, we have shown that for a massive spin-2 field $X=G$ universally coupled to matter fields, the decay rate  $H\to \gamma G$ is non vanishing, but quite suppressed by terms of order $m_G^2/\Lambda^2$, with $\Lambda$ the effective scale associated to the $H \gamma G$ coupling, and $m_G$ the particle mass. Same conclusions hold also for  higher spin fields, effectively coupled to the SM fields, whose contribution is also expected to be strongly  suppressed by the mass.
In conclusion, the potential measurement of the monochromatic photon signature in the $H\to \gamma X$ decay, with $X$ an invisible (light) dark particle $X$, would in practice uniquely identify $X$ as a dark photon, opening the way to the discovery of this particle as a portal to the dark sector.

All the above features promote the $H\to \gamma \DP$ decay to a golden channel for the dark photon discovery in both massless and massive scenarios. 

\vskip 1 cm
\noindent
{\bf \large Acknowledgments}

\noindent
We thank Damiano Anselmi, Matti Heikinheimo, and  Martti Raidal for inspiring discussions.


\end{document}